\documentclass{aa}  

\usepackage{txfonts,textcomp}
\usepackage{balance}
\usepackage{comment}
\usepackage{graphicx}
\usepackage[dvipsnames]{xcolor}
\usepackage{txfonts}
\usepackage{amsmath}
\usepackage[flushleft]{threeparttable}
\usepackage[normalem]{ulem}

\usepackage[breaklinks, colorlinks, citecolor=blue, urlcolor = blue]{hyperref}

\setcitestyle{citesep={,}}

\makeatletter
\renewcommand*\aa@pageof{, page \thepage{} of \pageref*{LastPage}}
\makeatother

\begin{document} 

   \title{Tormund's return: Hints of quasi-periodic eruption features from a recent optical tidal disruption event}
   \author{E. Quintin
          \inst{1}\thanks{erwan.quintin@irap.omp.eu},
          N. A. Webb\inst{1},
          S. Guillot\inst{1},
          G. Miniutti\inst{2},
          E. S. Kammoun\inst{1,3},
          M. Giustini\inst{2},
          R. Arcodia\inst{4}\thanks{Einstein Fellow},
          G. Soucail\inst{1},
          N. Clerc\inst{1},
          R. Amato\inst{1},
          \and
          C. B. Markwardt\inst{5}
}

   \institute{IRAP, Université de Toulouse, CNRS, UPS, CNES, 9 Avenue du Colonel Roche, BP 44346, 31028 Toulouse Cedex 4, France
             \and 
             Centro de Astrobiología (CAB), CSIC-INTA, Camino Bajo del Castillo s/n, ESAC campus, E-28692, Villanueva de la Cañada, Madrid, Spain
             \and
             INAF -- Osservatorio Astrofisico di Arcetri, Largo Enrico Fermi 5, I-50125 Firenze, Italy
             \and
             MIT Kavli Institute for Astrophysics and Space Research, 70 Vassar Street, Cambridge, MA 02139, USA
             \and 
             Astrophysics Science Division, NASA Goddard Space Flight Center, Greenbelt, MD 20771, USA
             }

   \date{}
    \titlerunning{Tormund's return}
    \authorrunning{Quintin et al.}
 
  \abstract
   {  Quasi-periodic eruptions (QPEs) are repeating thermal X-ray bursts associated with accreting massive black holes, the precise underlying physical mechanisms of which are still unclear.} 
   {  We present a new candidate QPE source, AT~2019vcb (nicknamed Tormund by the Zwicky Transient Facility collaboration), which was found during an archival search for QPEs in the \textit{XMM-Newton} archive. It was first discovered in 2019 as an optical tidal disruption event (TDE) at $z=0.088$, and its X-ray follow-up exhibited QPE-like properties. Our goals are to verify its robustness as QPE candidate and to investigate its properties to improve our understanding of QPEs.}
   {  We performed a detailed study of the X-ray spectral behaviour of this source over the course of the \textit{XMM-Newton} archival observation. We also report on recent \textit{Swift} and \textit{NICER} follow-up observations to constrain the source's current activity and overall lifetime, as well as an optical spectral follow-up.}
   {  The first two \textit{Swift} detections and the first half of the 30 ks \textit{XMM-Newton} exposure of Tormund displayed a decaying thermal emission typical of an X-ray TDE. However, the second half of the exposure showed a dramatic rise in temperature (from $53.5^{+9.2}_{-7.7}$ eV to $113.8^{+2.9}_{-2.7}$ eV) and 0.2--2~keV luminosity (from $3.2^{+1.6}_{-1.0}\times10^{42}$ erg~s$^{-1}$ to $1.19^{+0.05}_{-0.05}\times10^{44}$ erg~s$^{-1}$) over $\sim15$ ks. The late-time \textit{NICER} follow-up indicates that the source is still X-ray bright more than three years after the initial optical TDE.}
   {Although only a rise phase was observed, Tormund's strong similarities with a known QPE source (eRO-QPE1) and the impossibility to simultaneously account for all observational features with alternative interpretations allow us to classify Tormund as a candidate QPE. If confirmed as a QPE, it would further strengthen the observational link between TDEs and QPEs. It is also the first QPE candidate for which an associated optical TDE was directly observed, constraining the formation time of QPEs.}

   \maketitle
%

\section{Introduction}
\label{sec:Intro}

The X-ray transient sky is rich in complex, rare, and still puzzling phenomena. One of the latest additions to the family of rare X-ray transients are quasi-periodic eruptions (QPEs), first discovered in 2019 \citep{miniutti_nine-hour_2019}. These sources are characterised by intense bursts of soft X-rays, repeating every few hours, showing thermal emission with temperatures of $\sim50$~eV in quiescence, and reaching $\sim$100~eV at the peak. 
To date, only four bona fide QPE sources are known: \object{GSN 069} \citep{miniutti_nine-hour_2019}, \object{RX J1301.9+2747} \citep{giustini_x-ray_2020}, \object{eRO-QPE1} and \object{eRO-QPE2} \citep{arcodia_x-ray_2021}, along with one additional strong candidate, \object{XMMSL1 J024916.6-041244} \citep{chakraborty_possible_2021}. A sixth source, \object{2XMM J123103.2+110648}, has been suggested as a possible QPE source due to its optical and X-ray spectral and variability properties \citep{terashima_candidate_2012, miniutti_nine-hour_2019,webbe_variability_2023}, although its light curve is more reminiscent of quasi-periodic oscillations \citep[QPOs; e.g.][]{vaughan_bayesian_2010,reis_200-second_2012,gupta_possible_2018}.

In terms of timing properties, the duration of the bursts can vary, most being quite short ($<5$\,ks), with only eRO-QPE1 presenting a burst duration of $\sim25$\,ks. The recurrence time, which corresponds to the time between two consecutive bursts, ranges from 10~ks to 60~ks. However, \citet{arcodia_complex_2022} showed that this timescale does not necessarily remain constant for a given source. On a longer timescale, they are also transient in nature, with QPEs in GSN~069 being observed over the course of $\sim$1 year only, although the QPE lifetime may actually be longer \citep{miniutti_repeating_2023}. QPEs have been detected from relatively low-mass galaxies, around central black holes in the mass range of $10^5-10^7\,M_\odot$, with peak X-ray luminosities of $\approx 10^{42}-10^{43}\,\rm erg\,s^{-1}$. Two types of burst profile have been seen \citep{arcodia_complex_2022}: GSN~069 and eRO-QPE2 display isolated and regularly spaced peaks \citep{miniutti_nine-hour_2019,arcodia_x-ray_2021}, while RX~J1301.9+2747 and eRO-QPE1 show a more complex temporal evolution and overlapping peaks \citep{giustini_x-ray_2020,arcodia_x-ray_2021}. Finally, QPEs seem to show an observational correlation with tidal disruption events \citep[TDEs,][]{1988Natur.333..523R,gezari_tidal_2021}. TDEs are the disruption of a star by a massive black hole due to the tidal forces of the central mass. The resulting stellar debris creates a temporary accretion disc around the super-massive black hole (SMBH), which leads to a transient outburst over several months up to a few years. Out of the five known QPEs, two show a link with past X-ray TDEs \citep{miniutti_nine-hour_2019,chakraborty_possible_2021}, which is unlikely to be a coincidence considering the rarity of TDEs \citep[rate of $\sim \rm 6\times10^{-5}\,~yr^{-1}\,galaxy^{-1}$,][]{van_velzen_optical-ultraviolet_2020}. 
Additionally, the host galaxy properties of all the QPE sources are akin to those favoured for TDEs in terms of central black hole mass \citep{wevers_host_2022} or their post-starbust nature \citep{french_tidal_2016, wevers_host_2022}, which increases the probability of a stellar interaction with the central SMBH.

While the precise mechanism responsible for the emergence of QPEs is not yet clear, several models have been suggested to explain their properties. Initially, radiation-pressure disc instabilities were proposed \citep{miniutti_nine-hour_2019}, but the asymmetry in some of eRO-QPE1 eruptions, as well as considerations on the viscous timescales of the accretion flow, disfavoured this explanation \citep{arcodia_x-ray_2021}. While some changes to the magnetisation and geometry of the accretion flow compared to standard radiation pressure instability might solve the timescale issues \citep{sniegowska_possible_2020,sniegowska_modified_2023, kaur_magnetically_2022, pan_disk_2022}, the asymmetry remains problematic. \citet{raj_disk_2021} suggested a model of disc-tearing instabilities triggered by Lense-Thirring precession, which would separate a misaligned disc into several independant rings, leading to shocks between them and temporary enhancements of the accretion rate on shorter timescales than the viscous one. Magnification of a binary SMBH through gravitational lensing was suggested \citep{ingram_self-lensing_2021}, but it is currently disfavoured because of the chromatic behaviour of known QPEs \citep{arcodia_complex_2022}. Most other models involve one or more bodies orbiting the central massive black hole. \citet{xian_x-ray_2021} explained QPEs by the collision of a stripped stellar core with an accretion disc, most likely consisting of the debris of the stellar envelope. This type of model implies a previous partial TDE, which has the advantage of being consistent with the observational correlation between QPEs and TDEs. \citet{metzger_interacting_2022} presented a model based on the interactions of two counter-orbiting, circular, extreme-mass-ratio inspiral (EMRI) systems, in which accretion from the Roche lobe overflow of the outer stellar companion is temporarily and periodically enhanced by the proximity of the second inner stellar companion. Finally, QPEs can also be explained by repeated tidal stripping of an orbiting white dwarf \citep{king_gsn_2020,zhao_quasi-periodic_2022,wang_model_2022, chen_milli-hertz_2022, king_quasi-periodic_2022}, most likely captured through the Hills mechanism \citep[ejection of a binary companion,][]{hills_hyper-velocity_1988, cufari_using_2022}. \citet{wang_model_2022} showed that, in this model, the initial tidal deformation of the inbound white dwarf heats and inflates its envelope, which can be accreted onto the SMBH and provoke what appears to be a TDE. Recently, and still in the context of a mass transfer scenario due to Roche lobe overflow, models explaining QPEs via shocks between the incoming streams or between the stream and the existing accretion flow have been proposed by \citet{krolik_quasi-periodic_2022} and \citet{lu_quasi-periodic_2022}.

Additional detections and observations of QPEs are necessary to discriminate between the models and understand the nature of QPEs. With this aim, and as part of an ongoing study on the systematic exploitation of multi-instrument X-ray archives (Quintin et al., in prep), we searched for new QPE candidates previously missed in archival data. We looked for short-term variable, soft X-ray sources for which the position matched the centre of galaxies present in the GLADE+ catalogue \citep{dalya_glade_2022}. A comparable data-mining work was performed by \citet{chakraborty_possible_2021} on the 4XMM catalogue \citep{webb_xmm-newton_2020}, in which they found one new QPE candidate. While they looked for characteristic quasi-periodic pulses in the short-term light curves of the X-ray sources, our search was more generic in terms of variability (see more details in Sect.\ref{sec:XMM}). This allowed us to detect a new QPE candidate, 4XMM~J123856.3+330957. 

The optical counterpart of this source, \object{AT 2019vcb}, was originally detected as a transient optical event by the Zwicky Transient Facility \citep[ZTF,][]{bellm_zwicky_2014} on November 15, 2019 (ZTF19acspeuw, nicknamed Tormund), with total magnitude (not host corrected) peaking at 17.79, 17.91, and 18.0 in the i, r, and g bands, respectively. Additionally, it was detected by ATLAS (ATLAS19bcyz, peak differential magnitude of 18.415 in the orange filter) and Gaia (Gaia19feb, peak differential magnitude 18.73   in the g band) a few days later. Its brightening of about 1 magnitude from archival levels in the g, r, and i bands, and its decay over about 100 days led to a classification as a TDE. As part of a monitoring of the long-term multi-wavelength behaviour of TDEs, the ZTF collaboration obtained optical and X-ray follow-ups of the source. The optical observation allowed for a spectrum to be measured about two months after the peak; the observation revealed a line-rich spectrum, consistent with a H+He TDE \citep[see Fig. 1 in ][]{hammerstein_final_2022}. The authors used the \texttt{MOSFiT} \citep{guillochon_mosfit_2018} and \texttt{TDEmass} \citep{ryu_measuring_2020} models to estimate the mass of the central black hole, $M_{\rm BH}^{\tt TDEmass}\approx6.5^{+2.4}_{-1.7}\times10^6\,\rm M_\odot,$ and $M_{\rm BH}^{\tt MOSFiT}\approx8.3^{+0.8}_{-0.7}\times10^7\,\rm M_\odot,$ respectively. The host galaxy was identified as being relatively low mass ($M_{Gal}\approx10^{9.49\pm0.06}\,M_\odot$, the lowest mass of the studied sample of that article) at redshift $z=0.088$. It presented a rest-frame u-r colour of $1.55\pm0.03$, the lowest of the studied sample, and was the second-youngest of the sample in terms of age of stellar population. The X-ray follow-ups consisted of two observations by the \textit{Neil Gehrels Swift Observatory} (hereafter \textit{Swift}) and one \textit{XMM-Newton} observation respectively 3.5, 5, and 6 months after the optical peak. The X-ray follow-ups revealed very soft, thermal emission; the \textit{XMM-Newton} observation in particular revealed a large short-term variability that is consistent with the rising phase of a long-duration QPE, akin to eRO-QPE1 \citep{arcodia_complex_2022}. 

In this paper, we provide a detailed study of the available data as well as new follow-up data (Sects.\,\ref{sec:DataRed}, \ref{sec:SpecAnalysis}, and \ref{sec:Results}). We then analyse the spectro-temporal behaviour of this source to confirm it as a strong QPE candidate and assess the constraints this new candidate puts on the QPE formation and emission mechanisms (Sect.\,\ref{sec:Discussion}).

\section{Search \& data reduction}
\label{sec:DataRed}
\begin{figure*}
    \centering
    \includegraphics[width=\textwidth]{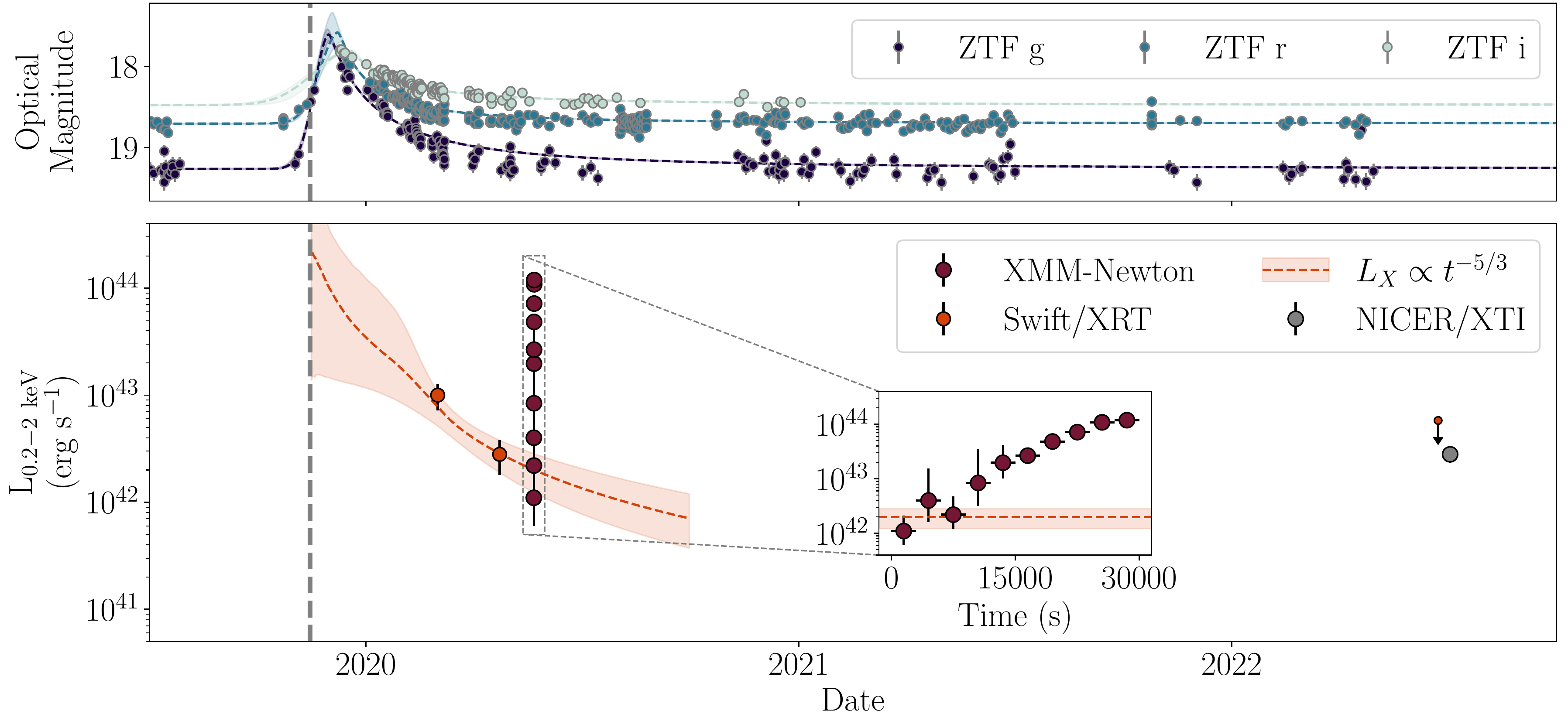}
    \caption{Multi-wavelength light curves of Tormund. The vertical grey dotted line corresponds to the date of detection of the optical transient by the ZTF collaboration. \textit{Top panel}: Optical g, r, and i magnitudes obtained from the ZTF catalogue. The optical magnitudes are not corrected for host galaxy emission. The dotted lines and shaded areas correspond to the posterior light curves obtained from fitting a Gaussian rise and power-law decay (see details in Sect.\,\ref{sec:DataRed}). For reference, the optical spectrum of the TDE (see top panel in Fig.\,\ref{fig:OpticalSpectrum}) was taken in February 2020 toward the end of the optical peak. \textit{Bottom panel}:\textit{Swift}, \textit{XMM-Newton}, and \textit{NICER} 0.2--2~keV luminosities. The orange dotted line and shaded area correspond to the estimated behaviour of an X-ray TDE decay phase following a $L_{X}\propto t^{-5/3}$ law, as extrapolated from the two \textit{Swift} data points (see Appendix \ref{sec:XTDE} for the precise method). The inset shows the short-term variability of the \textit{XMM-Newton} observation, with the quiescent state being consistent with the median value of the tail of the TDE, followed by a fast, large-amplitude burst.}
    \label{fig:all_light-curves}
\end{figure*}

The multi-instrument evolution of Tormund can be found in Fig.\,\ref{fig:all_light-curves}, and a summary of the X-ray observations is provided in Table\,\ref{tab:observations}.

\begin{table*}
\centering
\renewcommand{\arraystretch}{1.2}
\begin{tabular}{cccccc}
\hline \hline
Telescope & Instrument & ObsID & Date & Exposure \\ \hline \hline
Swift & XRT & 00013268001 & 01/03/2020 & 1.4ks\\ \hline       
Swift & XRT & 00013382001 & 23/04/2020 & 2.7ks\\ \hline
\textit{XMM-Newton} & EPIC-pn & 0871190301 & 22/05/2020 & 30ks\\
& EPIC-MOS1 &      &      & 32ks\\
& EPIC-MOS2 &      &      & 32ks\\
& OM/UVW1 &      &      & 7$\times$4.4ks\\ \hline
Swift & XRT & 00013268002 & 23/06/2022 & 1.6ks\\ \hline 
Swift & XRT & 00013268003 & 25/06/2022 & 1.5ks\\ \hline     
Swift & XRT & 00013268005 & 05/07/2022 & 2.1ks\\ \hline    
NICER & XTI & 5202870101 & 05/07/2022 & 5.1 ks\\ \hline
NICER & XTI & 5202870102 & 06/07/2022 & 8.3 ks\\ \hline
NICER & XTI & 5202870103 & 07/07/2022 & 4.8 ks\\ \hline
\end{tabular}
\caption{Summary of the X-ray data used in our study of Tormund. As a point of reference for the observation dates, the optical peak was detected by ZTF on November $15^{\text{th}}$, 2019. All exposures are effective exposures.}
\label{tab:observations}
\end{table*}

\subsection{\textit{XMM-Newton}}
\label{sec:XMM}

The source was found in the archival \textit{XMM-Newton} catalogue, 4XMM-DR11 \citep{webb_xmm-newton_2020}, as part of a larger project of data-mining the multi-instrument X-ray archives (Quintin et al., in prep.). We looked for short-term variable, soft, nuclear sources. To do this, we correlated the 4XMM-DR11 catalogue with a catalogue of galaxies, GLADE+ \citep{dalya_glade_2022}, which provides, among other things, position and distance estimates of about 23 million galaxies. We then selected the \textit{XMM-Newton} sources matching within 3$\sigma$ positional error bars with the centre of a GLADE+ galaxy, providing us with a list of about 40~000 nuclear X-ray-bright sources. We used the pre-computed variability estimate from the 4XMM-DR11 catalogue, \texttt{VAR\_FLAG} (which is a $\chi^{2}$ test on the short-term light-curve of the source for each observation) to select variable nuclear sources. Finally, we only kept the most spectrally soft sources by putting a threshold on the 0.2--2~keV to 2--12~keV fluxes hardness ratio, in the form of the condition $(F_{2-12keV}-F_{0.2-2keV})/(F_{2-12keV}+F_{0.2-2keV})<-0.9$. This allowed us to retrieve two known QPE sources (GSN 069, RX J1301.9+2747), a known possible QPE candidate (4XMM~J123103.2+110648), and the new QPE candidate, Tormund. Regarding the rest of the known QPEs, both eROSITA QPE sources were not yet publicly available in the 4XMM-DR11 catalogue, and the host galaxy of XMMSL1 J024916.6-041244 is not in GLADE+.

\begin{figure}
    \centering
    \includegraphics[width=\columnwidth]{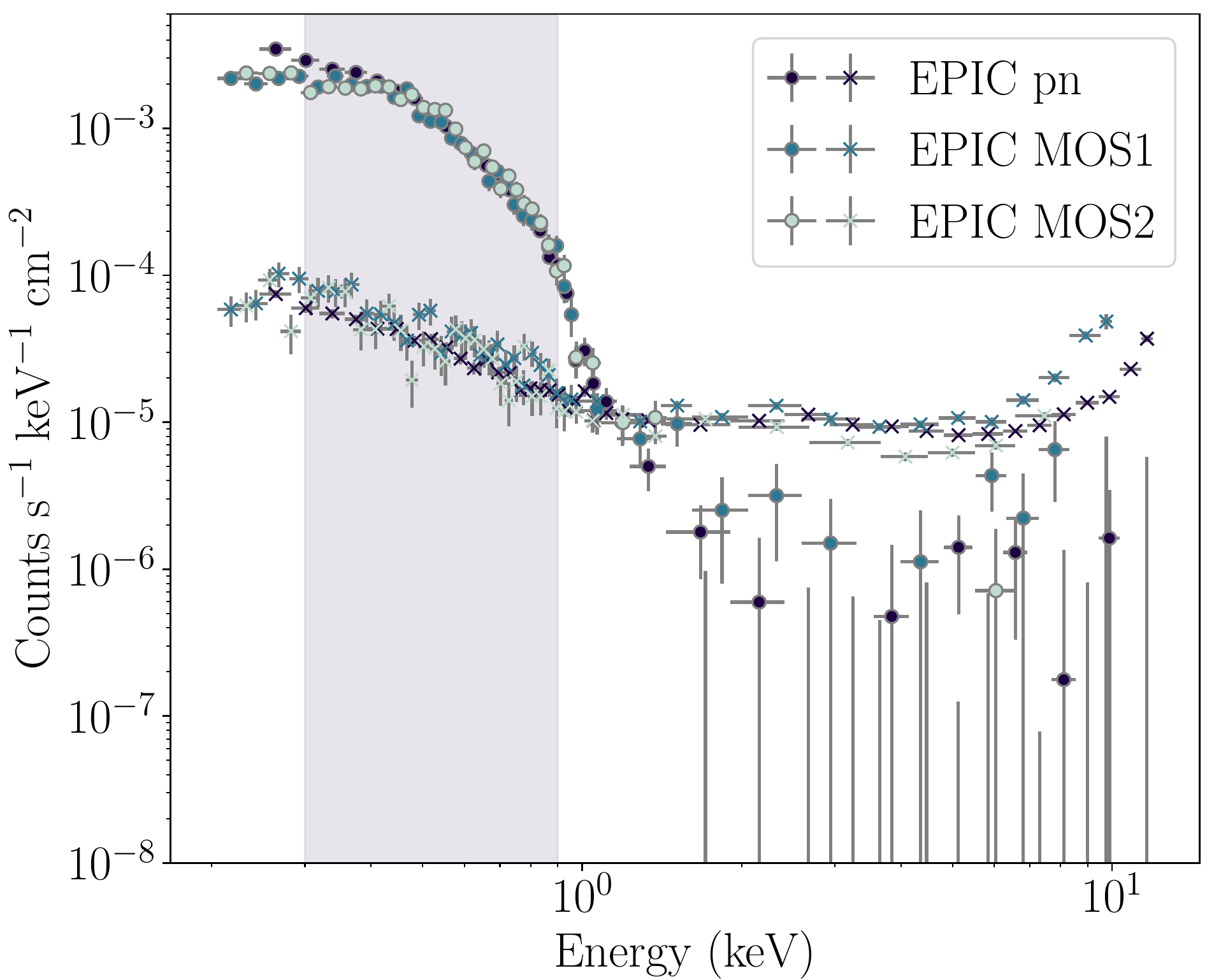}
    \caption{EPIC spectra of entire observation showing that the source (full circles) dominates the background (crosses) below 0.9 keV. Data below 0.3 keV are discarded to avoid calibration issues between EPIC instruments. The coloured area corresponds to the energy band we kept for the time-resolved spectral fitting.}
    \label{fig:EPICspectra}
\end{figure}

The archival \textit{XMM-Newton} observation (see Table\,\ref{tab:observations}) was about six months after the optically detected TDE peak. The data were reduced using the Science Analysis System (SAS) v.19.0.0, making simultaneous use of all EPIC instruments. The event lists were filtered for bad pixels and non-astrophysical patterns ($\leq$4 for pn and $\leq$12 for MOS 1 \& MOS 2).  A large, soft proton flare happened towards the last 5~ks of the observation, at the same time as the source reached its brightest state. According to the usual Good Time Interval (GTI) filtering method, based on an arbitrary threshold of the high energy ($\geq$10~keV) emission, the part of the light curve contemporaneous to the flare should be excluded, leading to the loss of the last 5~ks of the observation and half of the total detected photons. However, the extreme softness of the source allows us to mitigate the effect of this flaring background, which is overwhelmingly dominated by hard X-rays. This can be seen in Fig.\,\ref{fig:EPIClight-curves}, where the scaled background light curves (extracted for each instrument from a large empty nearby region on the same CCD) and the background-subtracted source light curves are shown, in both low (0.3--0.9~keV) and high (0.9--12~keV) energy bands. The background flare largely dominates the high-energy light curve (right panels), but not the low-energy one (left panels), where the contamination is well below the level of the background-subtracted source, even at the height of the flare. This confirms that we can keep the entire observation, including the last 5~ks, on the condition that we discard any data above about 0.9~keV. This energy threshold is further confirmed by the energy spectrum of the source and the background integrated over the entire duration of the observation, shown in Fig.\,\ref{fig:EPICspectra}, where the source dominates below 0.9~keV. We verified that this large high-energy background is independent of the position of the background extraction region, whether on the same CCD as the source or on another one. Additionally, for the first half of the observation, not subjected to the background flare, the count-rates of the source are relatively low ($\sim10^{-2}$ counts~s$^{-1}$ combined on all EPIC instruments), which leads to the source being above the background level only in this soft energy band as well. To be conservative and ensure the best signal-to-noise ratio for our data throughout the observation, and to avoid calibration issues between the EPIC instruments (see difference between the EPIC pn and MOS instruments in the 0.2--0.3~keV band in Fig.\,\ref{fig:EPICspectra}), we chose to limit our study of the \textit{XMM-Newton} data to the 0.3--0.9~keV band.

The Optical Monitor (OM) data, taken in fast mode, were extracted using the \texttt{omfchain} task; the source was the only one within the field of view of the Fast window. For each of the seven snapshots, we extracted the rates from the source and from a background region of 1.9".

Additionally, we reduced the data from the second \textit{XMM-Newton} observation of eRO-QPE1, ObsID 0861910301 \citep{arcodia_complex_2022}, in order to compare its properties with those of Tormund. We used the standard processing method and the usual GTI filtering method. For purposes of comparison with Tormund, we only kept the data in the 0.3--0.9~keV band.

\subsection{Swift}
A total of five \textit{Swift} observations were made of Tormund (see Table\,\ref{tab:observations}): two in January and February 2020 requested by the ZTF collaboration, and three in June 2022 as part of our follow-up study of this object, about two years after the \textit{XMM-Newton} observation. The data were processed using the automatic pipeline \citep{evans_methods_2009}. We retrieved the count rates or upper limits for all individual observations, as well as the combined spectrum for the first two observations, which were the only ones that led to detections. The 0.3--0.9 keV band was used for the \textit{Swift} data as well, as the softness of the emission prevented any detection at higher energies. 

\subsection{NICER}
A \textit{NICER} ToO was performed on Tormund in July 2022 (PI E. Quintin), a few days after the \textit{Swift} follow-up. A total of 20 ks was obtained in 12 consecutive exposures evenly sampled over 2.3 days, grouped in three successive daily ObsIDs. The data were processed using the \textit{NICER} Data Analysis Software \texttt{NICERDAS} v10 provided with \texttt{HEAsoft} v6.31, and calibration data v20221001.  Standard filtering criteria were used with the task \texttt{nicerl2}, with the exception of restriction on \texttt{COR\_SAX}$>1.5$\,GeV/c to exclude passages of \textit{NICER} in the polar horns of the Earth magnetic field causing  high background rates (particularly precipitating electrons) as well as restricting the undershoot rates with \texttt{underonly\_range='0-80'} to limit the effect of the low-energy noise peak below 0.4~keV (where a cold thermal component such as those of TDE is present).  The three available ObsIDs were combined into a single event file with \texttt{niextract-events}, followed by \texttt{ftmerge} to merge the \texttt{mkf} auxiliary files.  Finally, the 0.22--15~keV spectrum of the combined observations is generated with the tool \texttt{nicerl3-spec} using the 'SCORPEON'\footnote{\url{https://heasarc.gsfc.nasa.gov/lheasoft/ftools/headas/niscorpeon.html}} background model option. This generates scripts to perform spectral analyses of the source and background directly in Xspec. The SCORPEON background model provides both the measured spectral shapes of individual background components as well as a priori estimates of the normalisations of each component.  Within Xspec, it is possible to adjust the normalisations within a small range along with source parameters to better fit the measured spectrum.  Since the \textit{NICER} background is a broadband one, the use of the full 0.25-15 keV spectral fitting range improves the accuracy of the \textit{NICER} background estimate in the band of interest.  The SCORPEON model also has terms for known background features such as Solar Wind Charge Exchange (SWCX) emission lines, including partially ionised oxygen fluoresence. We also attempted to use the 3C50 background model \citep{remillard_empirical_2022}, but this model fails to account for the \ion{O}{VII} fluorescence emission line at 0.574~keV.

\subsection{Zwicky Transient Facility}
We retrieved the g, r, and i band light curves of Tormund from the ZTF archive \citep{masci_zwicky_2018}. These magnitudes are not corrected for the emission of the host galaxy. Each optical light curve was fitted with a Gaussian rise and a power-law decay \citep{van_velzen_seventeen_2021}:
$$L =L_{\mathrm{Quiescent}}+ L_{\mathrm{Peak}}\times \left\{
    \begin{array}{ll}
        e^{-(t-t_{\mathrm{peak}})^{2}/2\tau^{2}} & \mbox{if } t<t_{\mathrm{peak}}\\
        \big((t-t_{\mathrm{peak}}+t_{0})/t_{0}\big)^{-5/3} & \mbox{if } t\geq t_{\mathrm{peak.}}
    \end{array}
    \right.
$$
We used the \texttt{PyMC} framework \citep{salvatier_probabilistic_2016} with a Gaussian likelihood function and the \texttt{NUTS} sampler. We used 50 walkers on 3 000 steps, discarding the first 1 000. For each optical band, the median, 16$\textsuperscript{th,}$ and 84$\textsuperscript{th}$ percentiles of the associated posterior light curves were computed for each time step, which are shown by the dotted lines and shaded areas in the top panel in Fig.\,\ref{fig:all_light-curves}.

\subsection{MISTRAL}
The MISTRAL is a low-resolution spectrograph in the optical domain recently installed at the Cassegrain focus of the 1.93 metre telescope at Observatoire de Haute-Provence in France \citep{adami_mistral_2018}. Long-slit exposures with the blue grism and  covering the 400-800\,nm wavelength range at a resolution of $R \sim 1000$ took place on January 22, 2023 at around 05:00 UTC. The data set includes two exposures of 20\,min each in clear conditions (light cirrus and rare cloud passages). Additional data were acquired that are useful for data reduction (CCD biases, spectral flats with a Tungsten lamp, and wavelength calibration frames with HeAr lamps). An observation of the standard star Hiltner600 was carried out in the course of the run for flux calibration.

Data reduction was done with standard procedures using PYRAF\footnote{\url{https://iraf-community.github.io/pyraf.html}} for CCD correction and wavelength calibration. 2D images were cleaned from cosmic ray impacts and spectra were rebinned to 2 \AA / pixel. Finally, the spectrum of the galaxy was extracted and flux calibrated. The final spectrum is displayed in Fig.\,\ref{fig:OpticalSpectrum}, with the identification of the characteristic emission lines, redshifted at $z=0.0884$. 

Since most of the standard emission lines of star forming galaxies were detected in the spectrum, we computed their relative flux in order to locate the galaxy in the so-called BPT diagrams \citep{baldwin_classification_1981}. However, due to uncertainties in the flux calibration, any large-scale estimate of the flux distribution must be taken with caution.



\section{X-ray spectral analysis}
\label{sec:SpecAnalysis}

We performed a spectral-timing study of the eruption in the \textit{XMM-Newton} observation in two steps. At first, we extracted the combined EPIC background-subtracted light curves of the source in various energy bands  between 0.3 and 0.9 keV, in a similar fashion to \citet{arcodia_complex_2022}. The goal was to show the energy dependence of the start and rise times of the eruption; to estimate these parameters, we fitted the light-curves in each energy band with a simple burst model. In \citet{arcodia_complex_2022}, the model used was similar to those used for GRBs \citep{norris_long-lag_2005}, with an exponential rise and exponential decay. Here, the observation was not long enough to constrain any decay. The transition to the final plateau was also smoother than for eRO-QPE1. The model we used was thus simpler, with a Gaussian rise akin to that of TDE models \citep{van_velzen_first_2019} and then a constant plateau until the end of the observation. The fit was performed using the \texttt{curve\_fit} function from \texttt{SciPy} \citep{virtanen_scipy_2020}. To allow for a comparison with the parameters of eRO-QPE1, we computed the start and rise times of the eruption with the same method as \citet{arcodia_complex_2022}; the start time is the time where the count rate is $1/e^{3}$ the peak value. The rise time is then the difference between the peak of the Gaussian and this start time. 

To study the spectral-timing properties of the burst in depth, we  divided the observation into several time windows. For the first time window, lasting 1.5~ks, only the EPIC MOS1 and MOS2 instruments were turned on, so the low signal-to-noise ratio prevented a meaningful spectral study. For the rest of the observation, the three EPIC instruments were available, and the remaining exposure was sliced into a total of ten 3 ks windows. Each spectrum was binned to have one count per spectral bin. We performed systematic fitting of the ten time windows using \texttt{xspec} \citep{arnaud_xspec_1996} with the Cash statistic \citep{cash_parameter_1979} as implemented in \texttt{xspec} and abundances from \citet{wilms_absorption_2000}. 

We used two different spectral models. The simplest possible model is \texttt{tbabs$\times$zashift$\times$bbody}, for a single unabsorbed redshifted black body, with both temperature and normalisation of the black body being free parameters between time windows. The second, more complex model is \texttt{tbabs$\times$zashift$\times$(diskbb+bbody)}, for a dual component emission, with \texttt{diskbb} being linked between all time windows and corresponding to an underlying constant accretion disc emission, and \texttt{bbody} being free and corresponding to the eruptive feature. In both spectral models, the absorption was fixed at the Galactic value in the line of sight, $N_{\rm H} = 1.4\times 10^{20}$\,cm$^{-2}$ from the HI4PI collaboration \citep{ben_bekhti_hi4pi_2016}, as adding an extra intrinsic absorbing column density only resulted in upper limits, which were negligible compared to the Galactic value. We also performed fits of these models on the first four time windows combined, which corresponds to the quiescent state of the object. To quantify the goodness of the fits, we performed Monte Carlo simulations of the best-fit spectra for those slices with fewer than 25 counts per bin (i.e. the quiescent state and the first two eruption slices), for which the $C_{stat}$ alone would not be a good quantifier of the goodness of fit. We did not perform Monte Carlo simulations for the other spectra, where the higher signal allows for an interpretation of the Cash statistic in the Gaussian approximation.

To estimate the physical extent of the emitting region, we replaced each \texttt{bbody} component in each model with a \texttt{bbodyrad} component, which allowed us to retrieve the physical size of the emitting black body from the normalisation, given the distance and assuming a circular shape seen face-on. To compute the physical size of the emission region in a more precise way, we compared the evolution of the bolometric luminosity to the temperature (see more details in Sect. \ref{sec:Results}). These bolometric luminosities are derived from the best-fit normalisation of the black-body components. The 0.2--2~keV luminosities, used to compare to other QPEs, are computed by taking the bolometric luminosities and temperature and restricting it to the 0.2--2~keV band.

\section{Results}
\label{sec:Results}

The first X-ray data points of this source obtained after the optical TDE are two \textit{Swift} observations, respectively 3.5 and 5 months after the optical peak. They lead to two detections showing a very soft emission. The 0.3--0.9~keV count rates decreased by a factor of three over the 50 days separating these observations, from $(9.5\pm2.6)\times10^{-3}$~counts~s$^{-1}$ to $(3.5\pm1.2)\times10^{-3}$~counts~s$^{-1}$. The spectra are consistent with unabsorbed black bodies with respective rest-frame temperatures of $k_BT=76\pm15$~eV and $k_BT=130\pm70$~eV. The second temperature being poorly constrained due to the low counts, we combined both detections, assuming a constant temperature between them, yielding a better constrained combined temperature of $k_BT=76^{+12}_{-10}$~eV. Extrapolating to the 0.2--2~keV band using this temperature and the optically measured redshift of $z=0.088$, this translates into 0.2--2~keV unabsorbed rest-frame luminosities of $(1.0\pm0.28)\times10^{43}$~erg~s$^{-1}$ and $(2.8\pm1.0)\times10^{42}$~erg~s$^{-1,}$ respectively. No signs of intra-observation variability were detected, as both exposures were quite short (1.4 ks and 2.7 ks).

\begin{figure}
    \centering
    \includegraphics[width=\columnwidth]{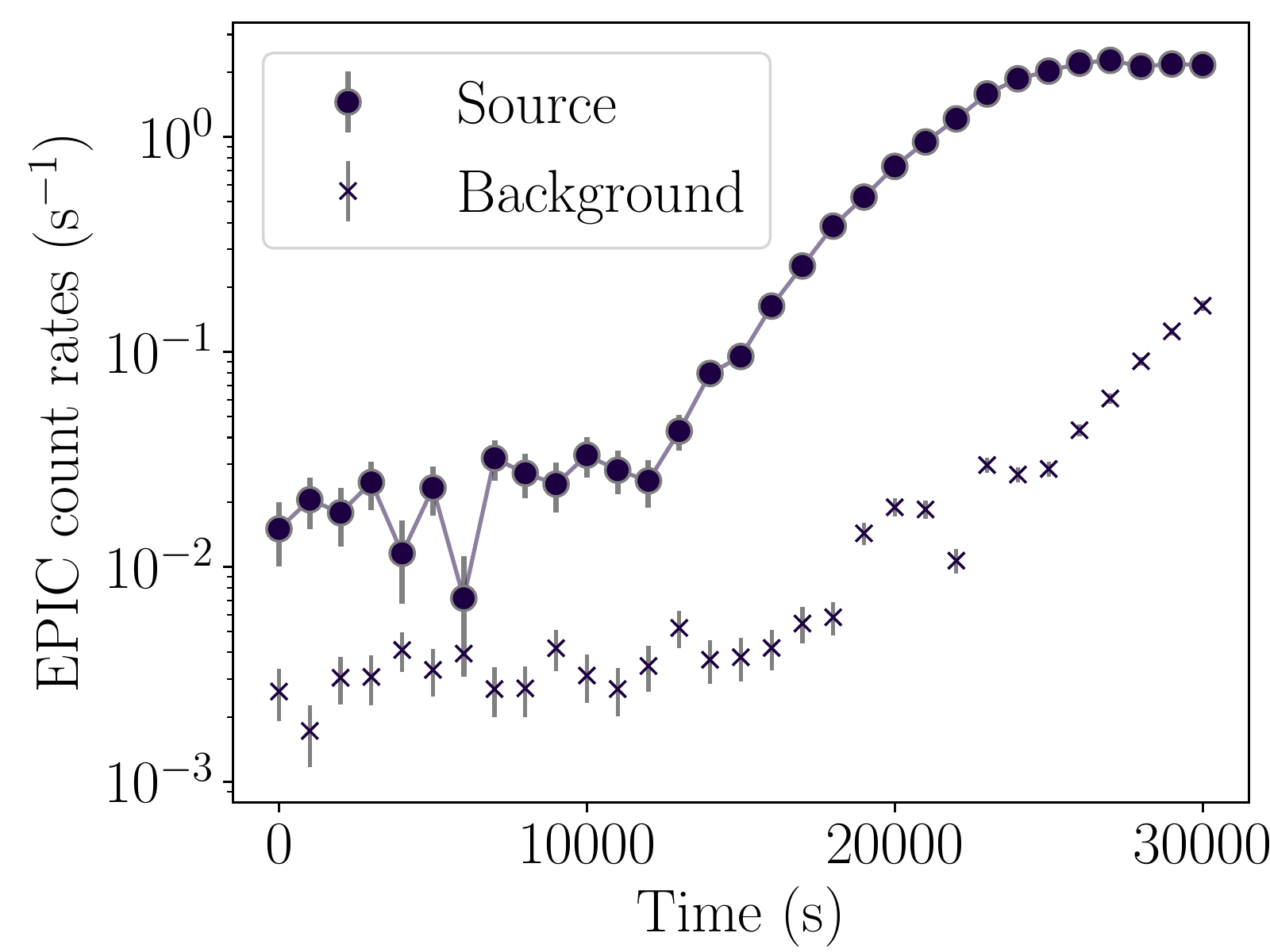}
    \caption{\textit{XMM-Newton} combined EPIC light curves of Tormund in 0.3--0.9~keV band, for both background-subtracted source and background, binned at 1000~s.}
    \label{fig:EPIC_Singlelight-curve}
\end{figure}

\begin{figure*}
    \centering
    \includegraphics[width=\textwidth]{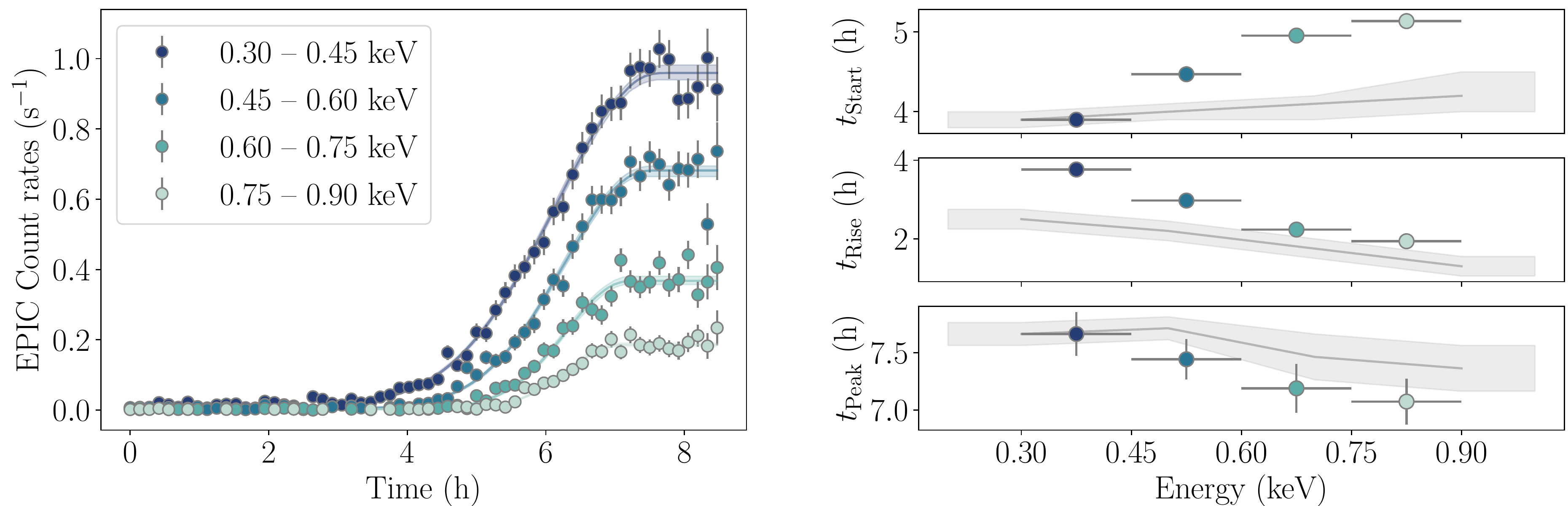}
    \caption{Short-term light curves and timing properties of the \textit{XMM-Newton} burst of Tormund in different energy bands. \textit{Left panel:} Combined background-subtracted EPIC light curves in different energy bands, binned at 500~s. Each light curve was fitted with a simple model of Gaussian rise between two plateau phases. The envelopes correspond to the 16$^{\rm th}$ and 84$^{\rm th}$ percentiles of the posteriors generated from the fitted parameters. \textit{Right panels:} Energy dependence of the fitted start time, rise time, and peak time of the burst. The start and peak times are expressed with respect to the start of the observation. The grey shaded areas correspond to the same parameters for eRO-QPE1 in \citet{arcodia_complex_2022}, with an offset for $t_\mathrm{Start}$ and $t_\mathrm{Peak}$ to overlap the curves -- it shows the similar energy-dependent behaviour between the sources (although the burst profile was exponential for eRO-QPE1).}
    \label{fig:light-curvesBands}
\end{figure*}

The \textit{XMM-Newton} observation, however, revealed a large short-term outburst in the soft X-rays a month after the second \textit{Swift} observation and six months after the optical TDE. Starting at about the middle of the exposure, the 0.3--0.9~keV combined EPIC count rates increased by a factor of $125^{+30}_{-20}$ from $(1.3\pm0.2)\times10^{-2}$~count~s$^{-1}$ to $1.7\pm0.1$~count~s$^{-1}$ (see Fig.\,\ref{fig:EPIC_Singlelight-curve}). This burst occurred over $\sim15$~ks. The last 3~ks of the observation showed a stabilisation of the count-rates, suggesting that this 15~ks rise time is indeed the characteristic rise time of the observed event, and not just limited by the end of the observation. The timescales of the burst are energy-dependent, as can be seen in the combined EPIC light curves in different energy bands shown in Fig.\,\ref{fig:light-curvesBands}. The fitted Gaussian burst profiles yield different values for the start time and rise time with increasing energies. The burst starts sooner for lower energies than higher energies ($\sim$1h delay between the start of the 0.3--0.45~keV and 0.75--0.9~keV bursts) and is faster at higher energies ($\sim$2h) than lower energies ($\sim$4h). The details are presented in Table\,\ref{tab:GaussianFitlight-curves}. These values and their energy dependences are similar to those obtained from eRO-QPE1 \citep{arcodia_complex_2022}. The light curves normalised to the peak values can be found in Fig. \ref{fig:NormalizedToPeak} showing the different rise times for each energy band \citep[similar to Fig. 2 of ][]{miniutti_nine-hour_2019}. An additional check on the necessity of energy-dependent parameters can be performed by simultaneously fitting the Gaussian burst profiles for each of the energy bands and tying the rise and peak times between them. This results in a significantly worse fit statistic, with the $\chi^{2}$/DoF increasing from the initial 99/98 to 360/104 when tying the time parameters between the energy bands. This validates the need for independent time parameters between the energy bands, that is the presence of energy-dependence in the rise profile.

\begin{figure}
    \centering
    \includegraphics[width=\columnwidth]{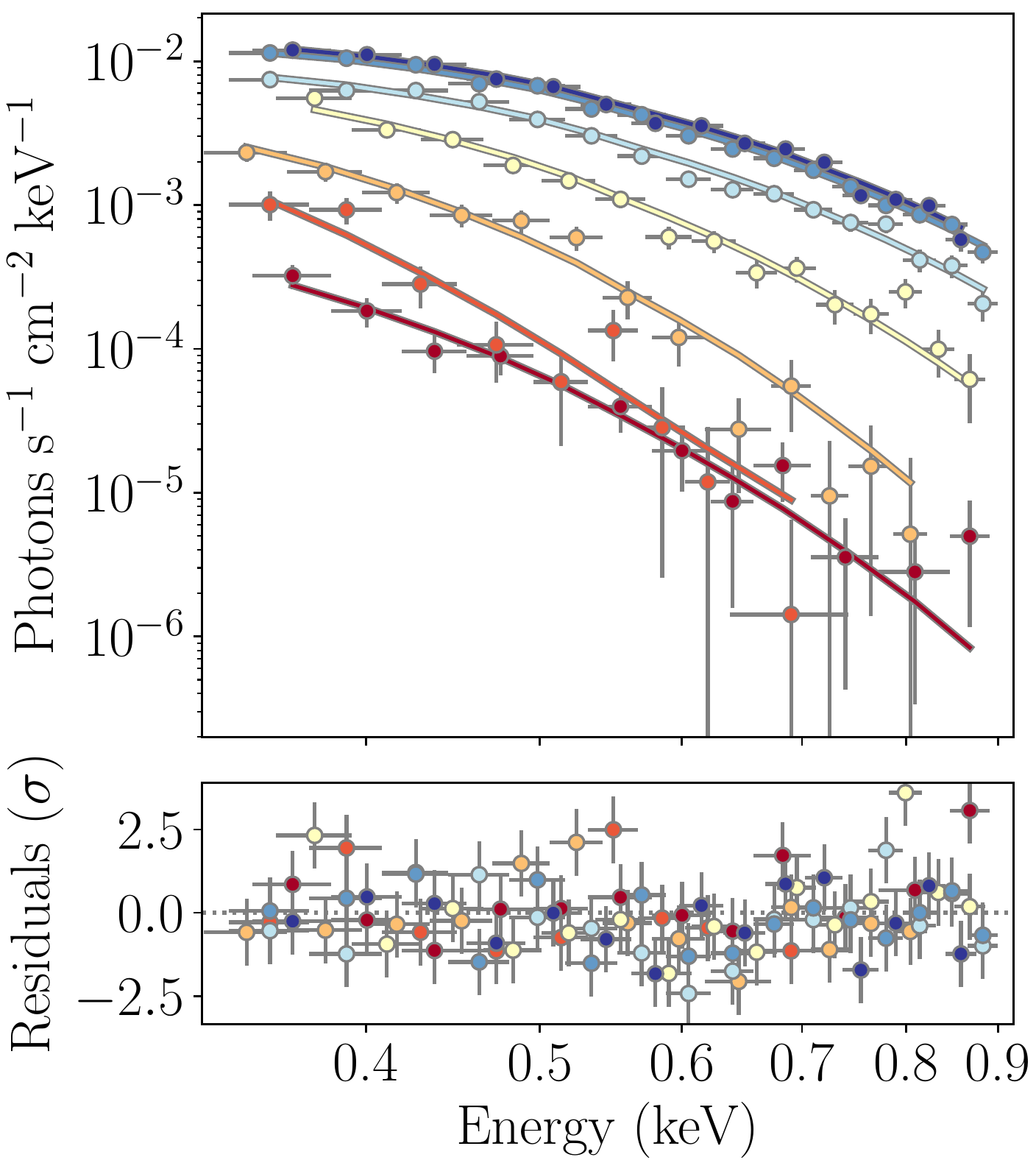}
    \caption{Time-resolved spectra of the \textit{XMM-Newton} observation fitted with a black body. The colours are used to identify individual time windows (see Fig.\,\ref{fig:SpectralTiming}).}
    \label{fig:Bbody_spectra}
\end{figure}

\begin{figure}
    \centering
    \includegraphics[width=\columnwidth]{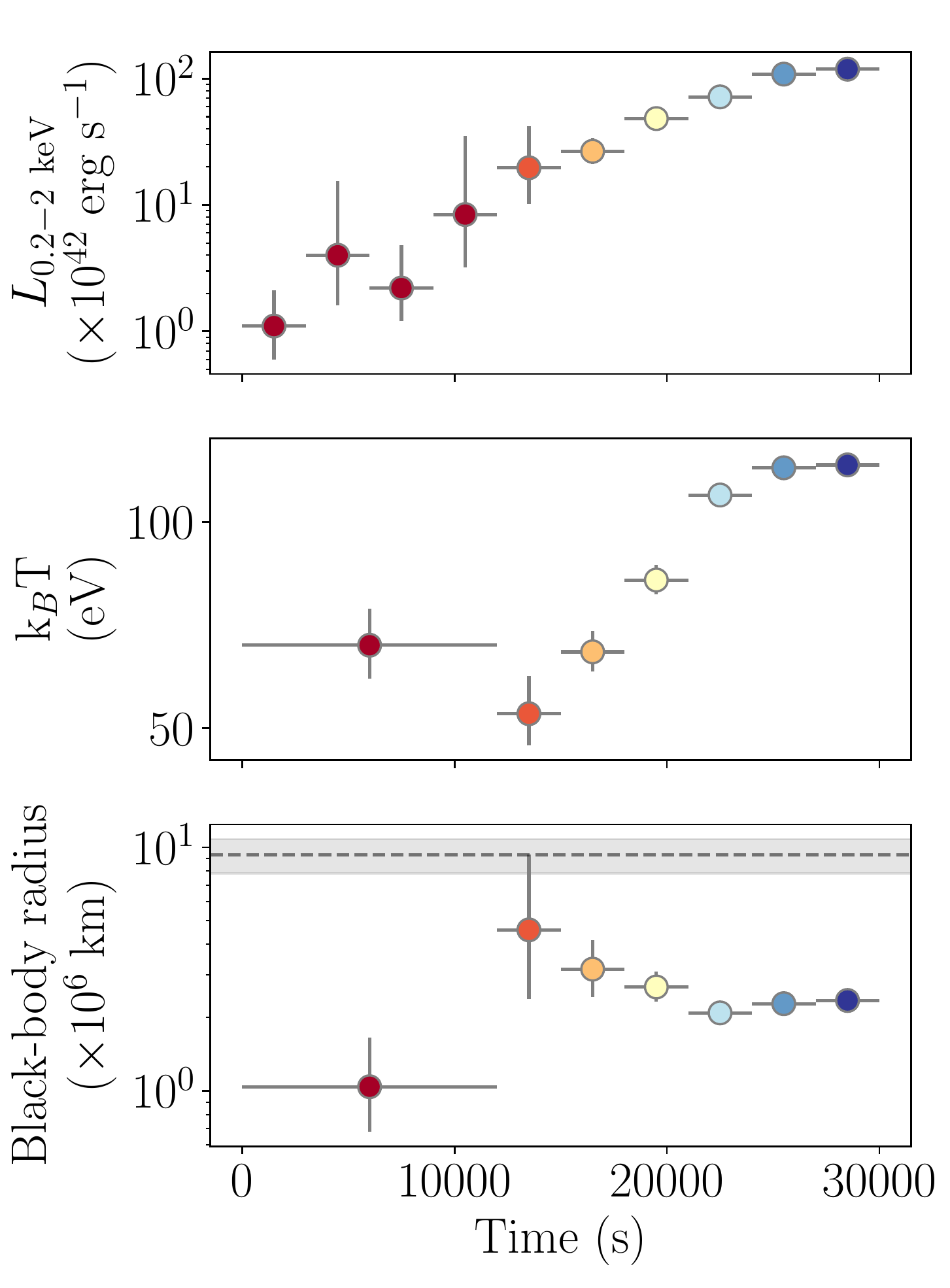}
    \caption{Results of spectral-timing study  of the \textit{XMM-Newton} observation for the \texttt{tbabs}$\times$\texttt{zashift}$\times$\texttt{bbody} model. \textit{Top panel:} Evolution of the 0.2--2.0 keV unabsorbed luminosity. \textit{Middle panel:} Evolution of the temperature of the black body. The quiescent state, corresponding to the first 12 ks, was combined. \textit{Bottom panel:} Evolution of the radius of the black body. The precise values and errors can be found in Table\,\ref{tab:fitbbody}. The grey dotted line and shaded area correspond to the gravitational radius value and errors inferred from the \texttt{TDEmass} black-hole mass estimate from \citet{hammerstein_final_2022}. This temporal evolution is available as an online movie.}
    \label{fig:SpectralTiming}
\end{figure}

To constrain the spectral-timing property of the burst more precisely, we then looked at the spectra fitted in time windows. The results of the fit are presented in Table\,\ref{tab:fitbbody} and Table\,\ref{tab:fitdiskbb} for both models. They both fit the data in a similar fashion, so only the fitted \texttt{bbody} model is shown in Fig.\,\ref{fig:Bbody_spectra}. The first model, a single black body, showed a steady increase of the temperature from 55 to 110~eV coincident with the increase in luminosity (see Fig.\,\ref{fig:SpectralTiming}). The first 12 ks, corresponding to the quiescent state and denoted as Time Window 0 in Table\,\ref{tab:fitbbody}, are marginally warmer at $70\pm8$~eV. For the second model, the \texttt{diskbb} component corresponds to the quiescent state, and the \texttt{bbody} component corresponds to the eruption feature. The two models are comparable in terms of flux and temperatures of their respective \texttt{bbody} components for the last five time windows. The fit statistics of both models are highly similar, so neither model is favoured when looking at the entirety of the observation. For the first slices with relatively low signal, the Monte Carlo estimation of goodness of fit confirmed the quality of the fit. We found percentages of the worst realisation of the fits of 12\% for the quiescent state and of 68\% and 4\% for the first two eruption slices, respectively, the latter being marginally acceptable.

\begin{figure}
    \centering
    \includegraphics[width=\columnwidth]{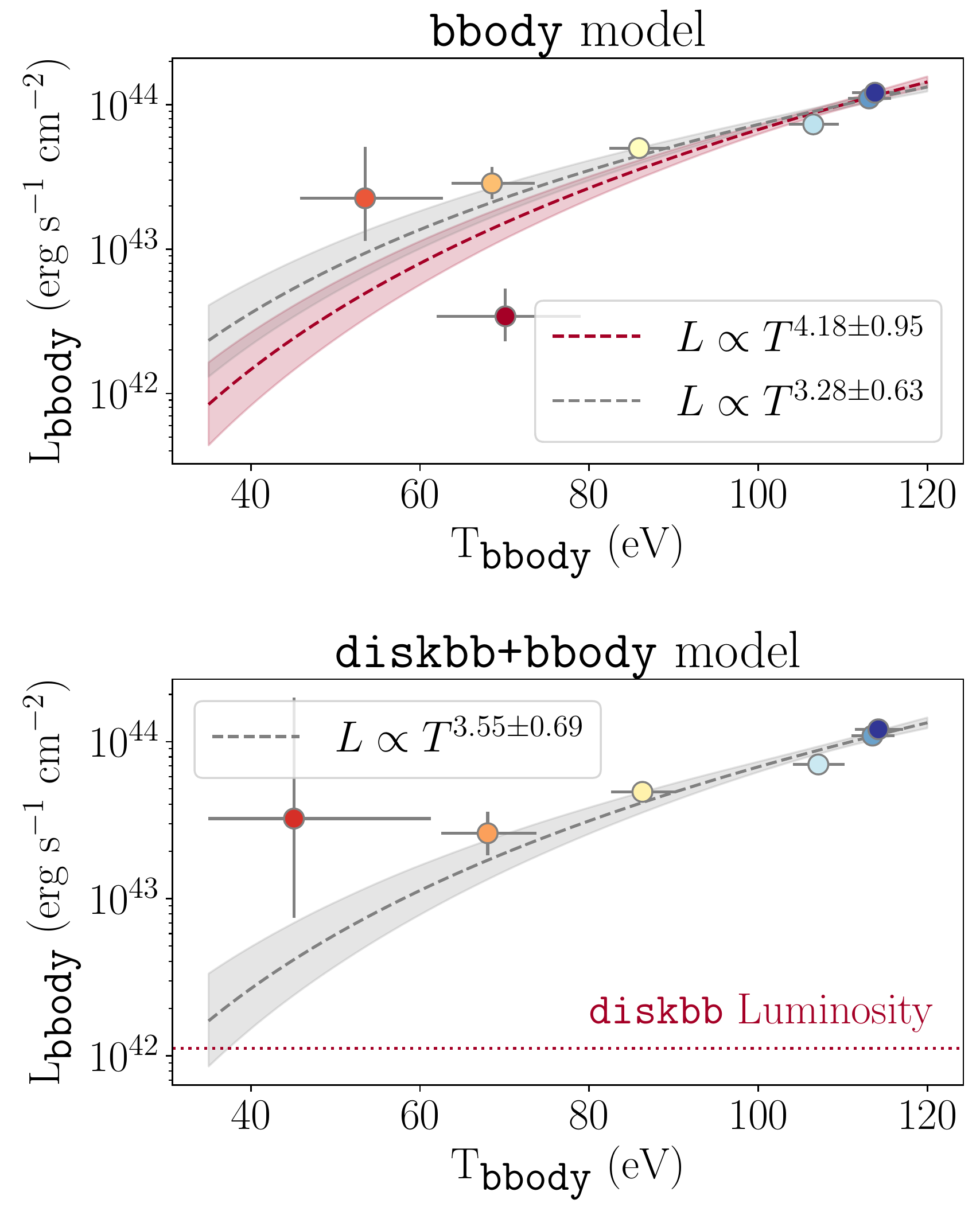}
    \caption{Evolution during the \textit{XMM-Newton} observation of the bolometric ${\tt bbody}$ luminosity compared to the temperature in both spectral models, fitted with a power law. The colours are used to identify individual time windows (see Fig.\,\ref{fig:SpectralTiming}). \textit{Top panel}: ${\tt TBabs\times zashift \times bbody}$ model. The red line corresponds to a fit using all the data points, with the quiescent state in dark red being a visible outlier. The grey line corresponds to the fit performed excluding the quiescent state. The shaded areas and the uncertainty on the power-law index correspond to a $1\sigma$ confidence level. \textit{Bottom panel}: ${\tt TBabs\times zashift \times (diskbb+bbody)}$ model. }
    \label{fig:EruptionFluxVersusT}
\end{figure}

The simultaneous evolution of the black body temperature from the \texttt{bbody} component in both models compared to the bolometric luminosity of this component is plotted in Fig.\,\ref{fig:EruptionFluxVersusT}. For both models, we fitted the luminosity evolution as a power-law function of the temperature, $L\propto T^{\alpha,}$ with $\alpha$ being a free parameter. For the first model (top panel), the quiescent state is represented as the outlying red dot, showing that it is marginally warmer but significantly fainter than the later time windows -- we performed the fit by including or excluding this point, which is shown, respectively, by a red or grey dotted line. In the case of the second model (bottom panel), the quiescent state is represented by the dotted line corresponding to the luminosity of the \texttt{diskbb} component. The fit parameters and statistics are shown in Table\,\ref{tab:LversusT}. All the fits are consistent at the $1\sigma$ level with $L\propto T^{4}$, but excluding the quiescent state for the first model greatly improves the fit statistic. Being consistent with $L\propto T^{4}$ means that the source can be interpreted as a pure black body of constant size heating up. We can thus compute the size of the emission region, by fitting the area $A$ in the Stefan-Boltzmann law, $L=A\sigma T^{4}$, and assuming a circular shape seen face-on. For the first model, we excluded the quiescent state from the fitting. Once again, the results are shown in Table\,\ref{tab:LversusT}. The first and second models result in consistent sizes of $\mathcal{R_{\texttt{bbody}}}=\left(1.30\pm0.05\right)\times 10^6$~km and $\mathcal{R_{\texttt{diskbb+bbody}}}=\left(1.27\pm0.04\right)\times 10^6$~km, respectively. The inferred radii are both consistent with radii fitted individually for each time window (see bottom panel in Fig.\,\ref{fig:SpectralTiming}), but they provide us with much tighter constraints. These radii are computed along the entire eruption. For the quiescent state only, the normalisation of the \texttt{bbody} model leads to a radius of $1.04^{+0.62}_{-0.36}\times10^6$~km, and the \texttt{diskbb} model leads to an inner radius of $0.75^{+0.53}_{-0.29}\times10^6$~km. For all the aforementioned radii, it is important to keep in mind that we assumed a face-on geometry and that no colour-correction for scattering within the emitting regions was taken into account -- both would mean that we underestimated the real physical size of the emitting regions by up to about an order of magnitude \citep{mummery_tidal_2021}. 

The \textit{XMM-Newton} observation was followed by a two-year gap in X-ray coverage, ended by our \textit{Swift} follow-up of the source. The three \textit{Swift} observations only lead to upper limits, constraining the total 0.3--0.9 keV count rate to be below $2.4\times10^{-3}$~counts~s$^{-1}$ at a $3\sigma$ level. Assuming a black-body spectrum at a temperature of 110~eV (justified by the following \textit{NICER} detection, see next paragraph), this leads to a 0.2--2~keV luminosity $3\sigma$ upper limit of $\sim6\times10^{42}$ erg~s$^{-1}$.

\begin{figure}
    \centering
    \includegraphics[width=\columnwidth]{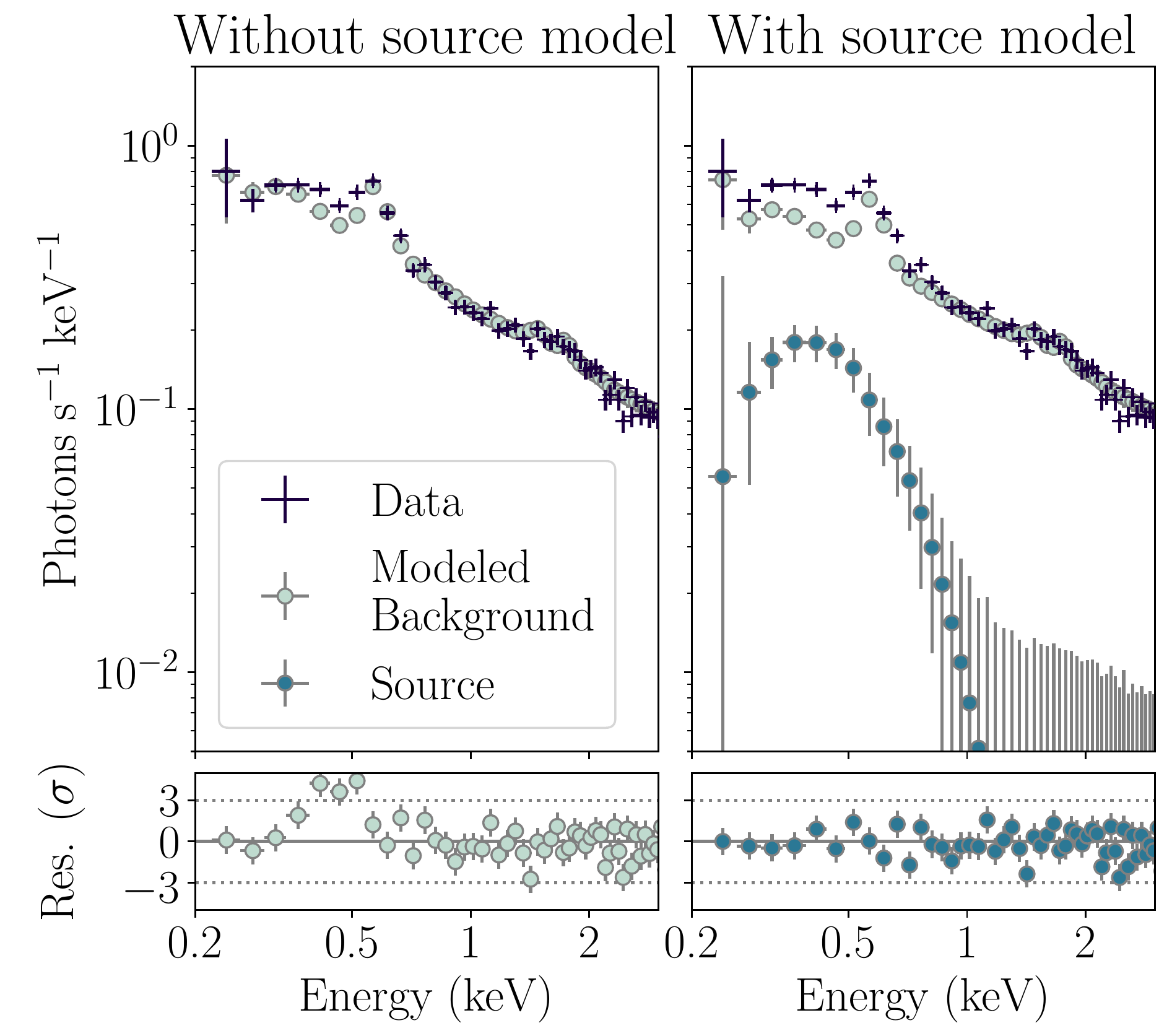}
    \caption{Combined \textit{NICER} best-fit background and model spectra and residuals of Tormund. \textit{Left panel:} Only background fitted model (SCORPEON), leading to significant residuals around $\sim0.5$~keV. \textit{Right panel:} Both background model and an additional black-body component are fitted. The fit is improved with better residuals.}
    \label{fig:NICER_spectrum}
\end{figure}

\begin{figure}
    \centering
    \includegraphics[width=\columnwidth]{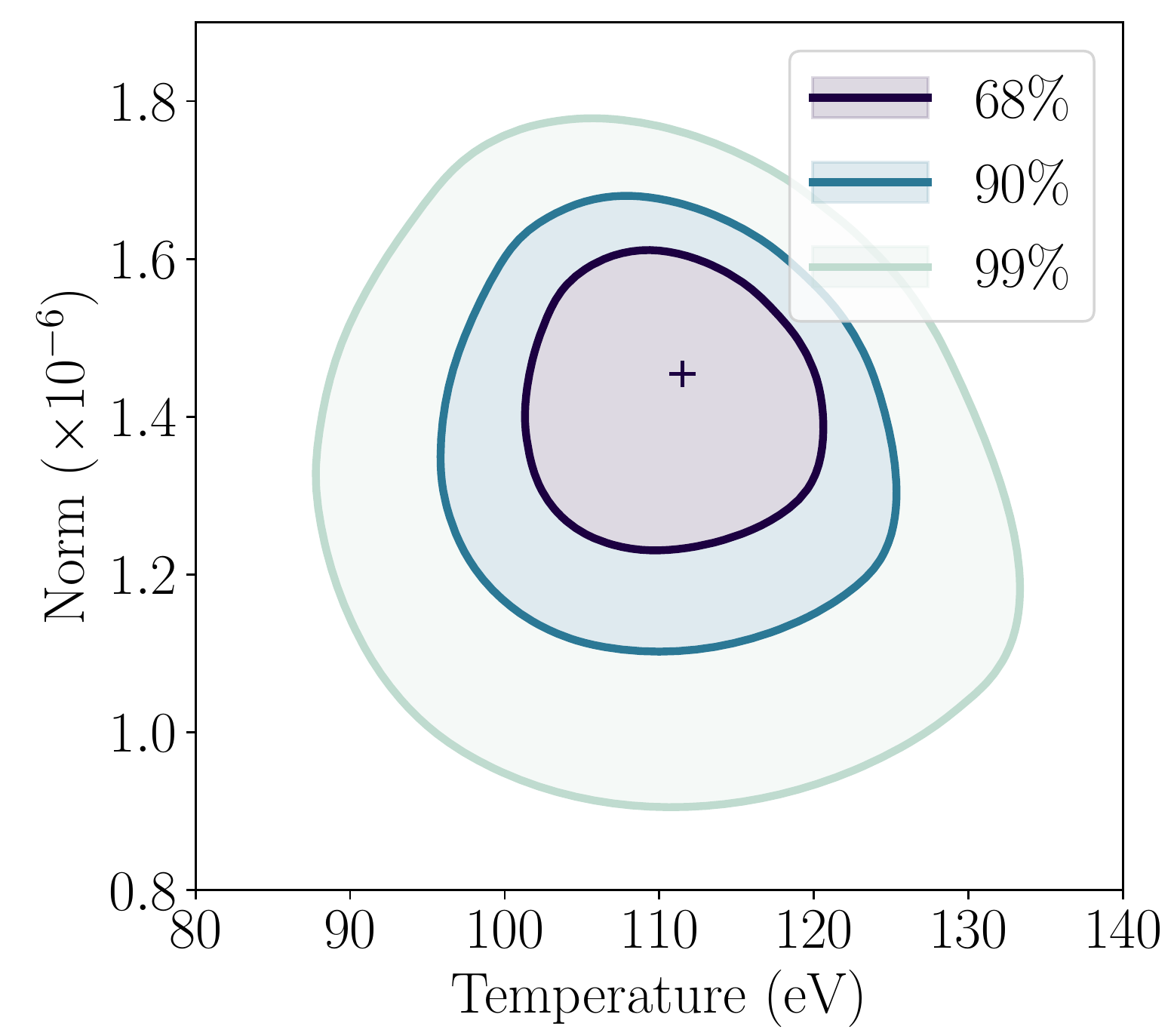}
    \caption{Contour plot of the spectral fit of the \textit{NICER} detection. The levels correspond to the $68\%$, 90\%, and 99\% confidence intervals, based on the difference in fit statistics to the best fit (cross). For visualisation purposes, the underlying 100$\times$100 grid was smoothed with a 2$\times$2 Gaussian blur. }
    \label{fig:NICER_contours}
\end{figure}

The \textit{NICER} follow-up, performed a week after our \textit{Swift} follow-up, led to further detections of soft emission from the source. As demonstrated in Fig.\,\ref{fig:NICER_spectrum}, fitting the spectrum with the SCORPEON model alone leads to significant and broad residuals near 0.5 keV, and we thus conclude that the source is detectable.  Adding a black-body component significantly improves the quality of the fit. All three individual snapshots were thus fitted with an unabsorbed black body, leading to similar but poorly constrained temperatures, $k_{B}T_{1}=105\pm57$~eV, $k_{B}T_{2}=107\pm40~$eV, $k_{B}T_{3}=103\pm50~$eV, and similar 0.2--2~keV luminosities of $L_{1}=(2.05^{+1.18}_{-1.15})\times10^{42}$~erg~s$^{-1}$, $L_{2}=(2.52 ^{+ 0.92 }_{- 0.94 })\times10^{42}$~erg~s$^{-1}$, and $L_{3}=(3.24 ^{+ 1.61 }_{- 1.63 })\times10^{42}$~erg~s$^{-1}$. The signal-to-noise ratio was too low to conclude on any intra-snapshot variability. The absence of strong sign of variability between snapshots, at least with an amplitude comparable to what was seen by \textit{XMM-Newton}, motivated the use of a combined \textit{NICER} fit. The combined \textit{NICER} data lead to a temperature of $111.5\pm19$ eV and a normalisation of $(1.45\pm0.25)\times10^{-6}$ (see Fig.\,\ref{fig:NICER_spectrum} and Fig.\,\ref{fig:NICER_contours}), corresponding to a 0.2--2 keV luminosity of $2.82 ^{+ 0.48 }_{- 0.48 }\times10^{42}$ erg~s$^{-1}$.

Finally, the MISTRAL follow-up of Tormund, performed about six months after the \textit{NICER} follow-up, provided us with a late-time optical spectrum to compare to the one obtained with the Keck observatory, during the decaying phase of the initial optical TDE \citep{hammerstein_final_2022}. The late-time optical spectrum (see Fig.\,\ref{fig:OpticalSpectrum}) revealed a strong H$\alpha$ line compared to the few other present lines, with log([NII]/H$\alpha)=-0.70\pm0.05$, log([SII]/H$\alpha)=-0.32\pm0.1$, and an equivalent width for the H$\alpha$ line of $-27\pm3$. The OI line is not detected, with log([OI]/H$\alpha$)$<-1.6$. Comparing this new spectrum to the initial TDE spectrum, the broad He II line and the broad H$\alpha$ feature, both directly linked to the TDE \citep{gezari_tidal_2021}, disappeared over the two years separating these observations. The continuum evolved as well, with a slightly redder emission in the late-time observation. The H$\beta$ line is quite dim, with log([OIII]/H$\beta)=0.33\pm0.1$ and H$\alpha$/H$\beta$=6.5$\pm2$. Large uncertainties remain in the H$\alpha$/H$\beta$ because of poor flux calibration of MISTRAL data due to relatively low exposure times. 
Using a Calzetti extinction law  with log(H$\alpha$/H$\beta$)~=~0.46+0.44 $E(B-V)$ \citep{calzetti_dust_2001,osterbrock_astrophysics_2006}, we find $E(B-V)=0.8\pm0.3$. This is indeed large, but consistent with the value found from photometric SED fitting in \citet{hammerstein_final_2022}, which was $E(B-V)$=0.67$\pm0.2$. Using the ratios of neighbouring lines that are less affected by flux calibration, the position of the source in the BPT diagrams is depicted in Fig.\,\ref{fig:BPT}. The source falls in the HII region of the [OIII]/H$\beta$ versus [NII]/H$\alpha$ diagram, at the crossing point of the three regions in the [OIII]/H$\beta$ versus [SII]/H$\alpha$ diagram, and the upper limit on the [OI] line makes it fall in the HII region of the [OIII]/H$\beta$ versus [OI]/H$\alpha$ diagram. Accounting for stellar absorption of the H$\beta$ line, for instance by fitting a template galactic component \citep{wevers_host_2022}, would lead to a larger H$\beta$ feature, so a smaller [OIII]/H$\beta$, driving the source even further down in the HII regions of all BPT diagrams. The W$_{\mathrm{H}\alpha}$ versus [NII]/H$\alpha$ (WHAN) diagram \citep{cid_fernandes_comprehensive_2011} leads to a classification as a star forming galaxy (Fig.\,\ref{fig:Optical_WHAN}). Both the BPT and WHAN emission lines diagnostics concur in excluding the presence of an AGN in Tormund's host galaxy.

\section{Discussion}
\label{sec:Discussion}
\subsection{Nature of the source}

The first step of this study was to confirm the identification of this source as a QPE candidate by excluding any other possible interpretation. First, we ruled out the possibility of a non-astrophysical source. Whilst a high-energy flare was present in the data due to a soft proton flare, it was clear that the soft variability we detected was indeed related to the astrophysical source (see Figures \ref{fig:EPIClight-curves} and \ref{fig:EPICspectra}). The spectral softness of the source is inconsistent with the expected hardness of the background flare, and limiting our study of this observation to the soft emission below 0.9~keV allows us to exclude the possibility that this variability is due to a proton flare. We stress that any conclusion drawn from the last $\sim 5\,\rm ks$ of the observation is dependent on the acceptance of this specific method, as the standard approach would simply discard this data altogether.

Secondly, we excluded any other astrophysical interpretations. 
Tormund having been observed by \textit{XMM-Newton} as part of a follow-up study of the optically-detected TDE, it can be assumed that the observed X-ray flare originates directly from the X-ray TDE itself. There are two ways to explain such a large and fast flux increase within 15~ks for a TDE: either the \textit{XMM-Newton} observation caught the TDE right at the time it started to become X-ray bright, during its rising phase delayed with respect to the optical peak \citep[as was seen, for instance, in the case of OGLE16aaa,][]{kajava_rapid_2020}; or, it was a late flare from the decaying TDE. Both scenarios struggle to explain all the observational properties of the source. For the first scenario, the main issue arises from the two previous \textit{Swift}/XRT detections, one and three months before the \textit{XMM-Newton} observation. They were already consistent with the decay phase of a TDE, in that they showed a very soft X-ray emission declining over time. We can additionally extrapolate the decay after the two \textit{Swift} observations assuming the standard $L \propto t^{-5/3}$ evolution of TDE bolometric luminosity over time \citep[e.g.][]{gezari_tidal_2021}. Since we cannot constrain the temperature evolution between the two \textit{Swift} observations, we assume a constant temperature for simplicity -- this translates into a flux evolution of $F \propto t^{-5/3}$. We can then compare the quiescent level of the \textit{XMM-Newton} observation to the expected flux of the decayed X-ray TDE, in the same manner as \citet{miniutti_repeating_2023} for GSN 069. We find that the \textit{XMM-Newton} quiescent luminosity is consistent with the expected rate from the general $F_{X} \propto t^{-5/3}$ evolution over time (see orange dotted line and shaded area in Fig.\,\ref{fig:all_light-curves}, and Appendix \ref{sec:XTDE} for details). The X-ray decay between the Swift detections and the \textit{XMM-Newton} quiescent state is thus consistent with what would be expected in a TDE. This strengthens the idea that the X-ray counterpart to Tormund was already behaving like an X-ray TDE during the two \textit{Swift} observations and prior to the \textit{XMM-Newton} short-term outburst. 
An additional point can be made about the improbability of observing the TDE in its rise by chance. Indeed, the X-ray counterparts to optical TDEs have sometimes been detected with significant delays of several months \citep{gezari_tidal_2021}. However, the \textit{XMM-Newton} exposure was not triggered on a particular re-brightening event, but rather a standard follow-up six months after the optical TDE. We roughly quantified the probability of detecting serendipitously, during a randomly-timed follow-up, the start of the rise of the X-ray TDE. We conservatively assumed a uniform optical-to-X-ray delay distribution of up to one year based on the properties of the few objects identified so far (about 10). Detecting the delayed X-ray TDE within a 30~ks exposure taken at a random time would then have an $\sim0.1\%$ chance of happening, making this serendipitous detection unlikely. Combined with the two prior \textit{Swift}/XRT detections, it thus excludes the first scenario, where the short-term variability we see is the rising phase of the X-ray TDE in itself.

A further explanation would be a late re-brightening from the already existing TDE. However, the amplitude is too extreme to be explained by this interpretation. During the \textit{XMM-Newton} observation, the 0.3-0.9 keV combined EPIC count rates rose by a factor of $\sim125$ in about 15 ks. This is not expected from short-term flares in X-ray TDE light-curves, with smaller amplitudes of approximately a factor of 5 and longer timescales of a few days in the sources detected so far \citep{wevers_evidence_2019, van_velzen_seventeen_2021, yao_tidal_2022}. Large-amplitude re-brightenings have been observed in X-ray TDEs \citep[see, for recent examples,][]{malyali_erasst_2023,malyali_rebrightening_2023}, but with larger characteristic rise times of several days rather than a few hours.
We can also exclude a supernova in the galactic nucleus, since the observed luminosities are too high \citep[typically $10^{35}$--$10^{41}$~erg~s$^{-1}$,][]{dwarkadas_what_2012}, and a prior optical counterpart for the supernova, independent of the TDE, would be expected -- which was not seen.

\begin{figure*}
    \centering
    \includegraphics[width=\textwidth]{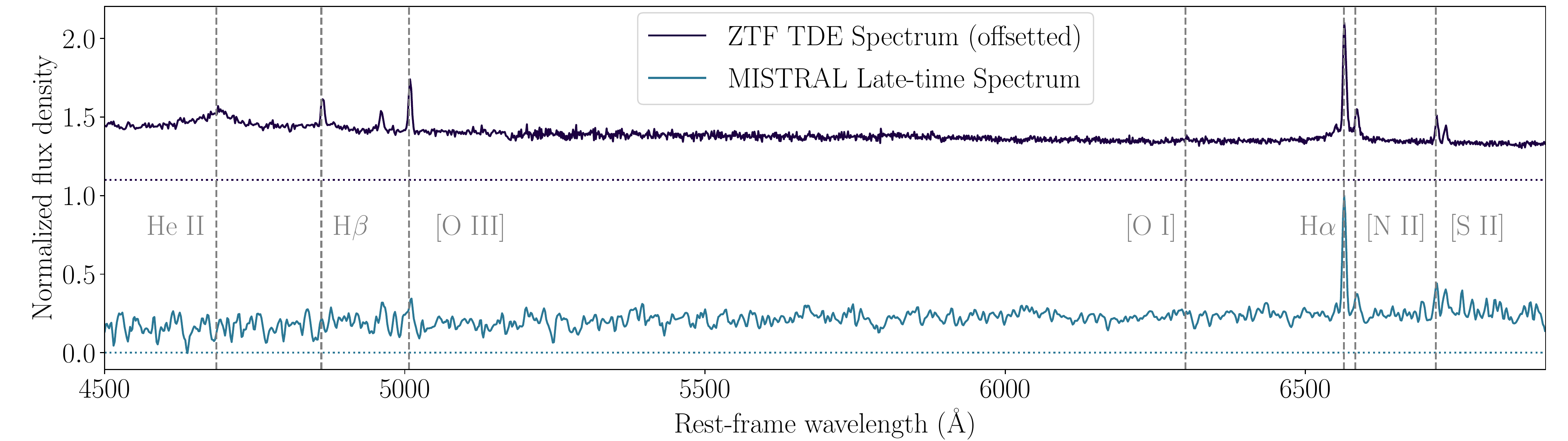}
    \caption{Background-subtracted optical spectra of Tormund, with data from the Keck follow-up performed by the ZTF collaboration during the initial TDE decay \citep{hammerstein_final_2022} at the top and from the late-time MISTRAL follow-up at the bottom. Both spectra were normalised to their respective maximum flux density values, and the ZTF spectrum was offset by +1 for visualisation purposes. One can notice 
    the disappearance of both the broad He II line and the broad H$\alpha$ feature.}
    \label{fig:OpticalSpectrum}
\end{figure*}

\begin{figure*}
    \centering
    \includegraphics[width=\textwidth]{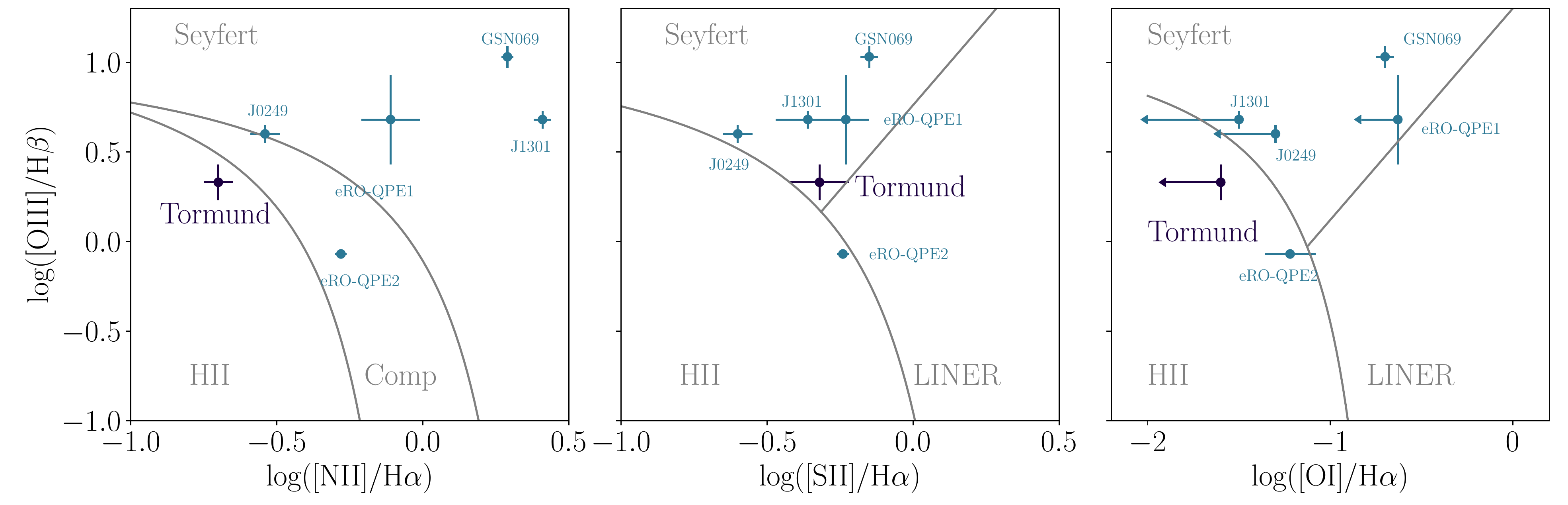}
    \caption{BPT diagram of QPE sources, including Tormund, indicating that its galaxy most likely does not host an AGN. For the other QPE sources, the data were taken from \citet{wevers_host_2022}. The relative weakness of the H$\beta$ line could most likely be explained by stellar absorption \citep{dewangan_active_2000, wevers_host_2022}, which is expected to be significant in host galaxies of TDEs, which are predominantly post-starbust \citep{french_tidal_2016}. Accounting for this absorption, for instance by fitting a template galactic component \citep{wevers_host_2022}, would lead to a larger H$\beta$ feature, and thus a smaller [OIII]/H$\beta$, driving the source even further down in the HII regions of all BPT diagrams.}
    \label{fig:BPT}
\end{figure*}

\begin{figure}
    \centering
    \includegraphics[width=\columnwidth]{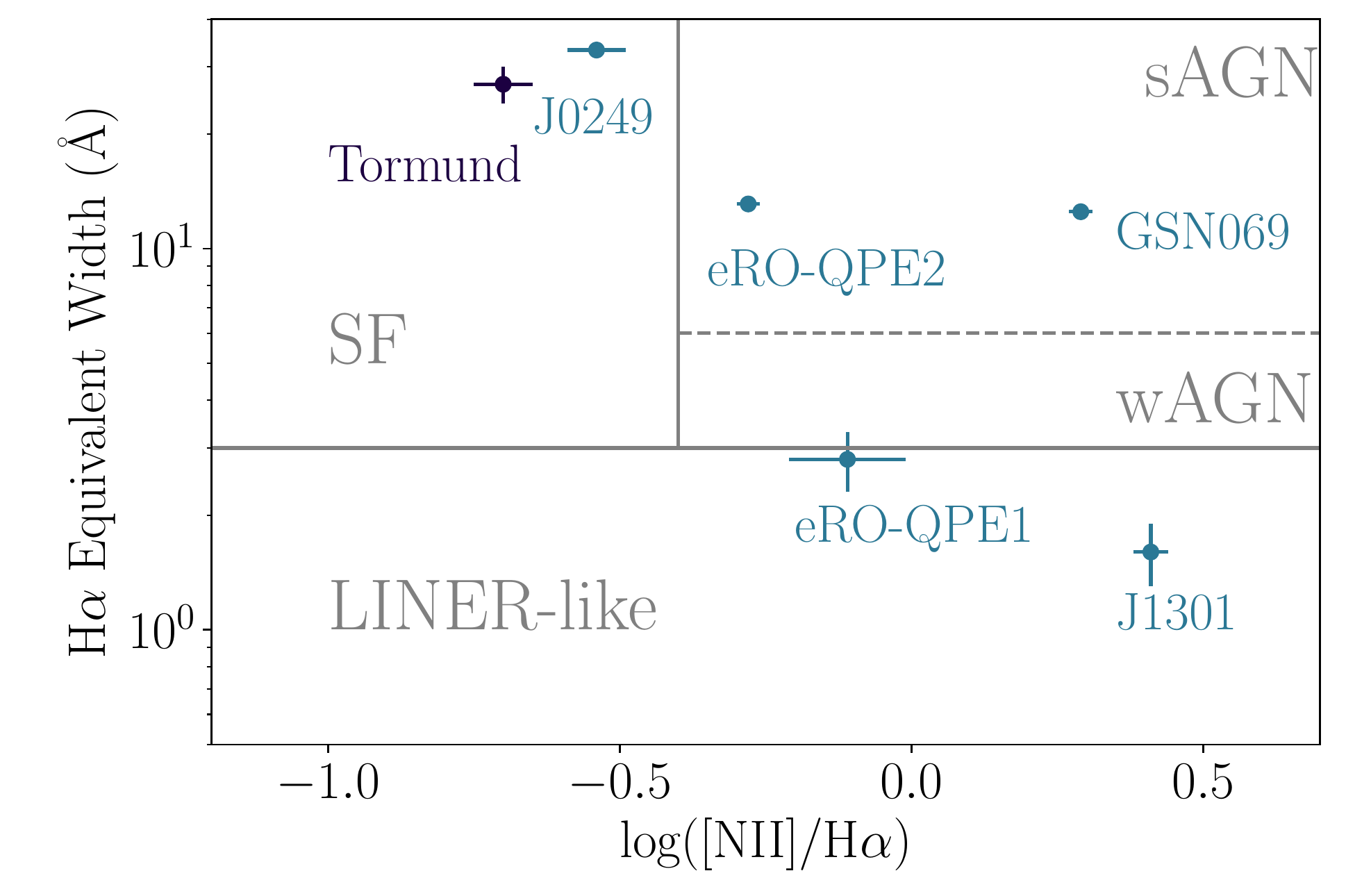}
    \caption{WHAN diagram of QPE sources, as taken from \citet{wevers_host_2022}, along with Tormund. This concurs with the BPT diagrams to conclude the absence of nuclear activity in Tormund's host galaxy.}
    \label{fig:Optical_WHAN}
\end{figure}

The only remaining astrophysical explanation would be an AGN flare. 
The strongest argument against this interpretation comes from the optical late-time spectrum (Fig.\,\ref{fig:OpticalSpectrum}). It showed weak emission lines apart from H$\alpha$, especially a very weak H$\beta$ line. Using both the BPT diagram \citep{baldwin_classification_1981} and the W$_{H\alpha}$ versus [NII]/H$\alpha$ (WHAN) diagram \citep{cid_fernandes_comprehensive_2011}, we can classify the galaxy as a star forming galaxy, with no sign of nuclear activity (see Fig.\,\ref{fig:BPT} and Fig.\,\ref{fig:Optical_WHAN}). This allows us to exclude the AGN interpretation.

\begin{figure}
    \centering
    \includegraphics[width=\columnwidth]{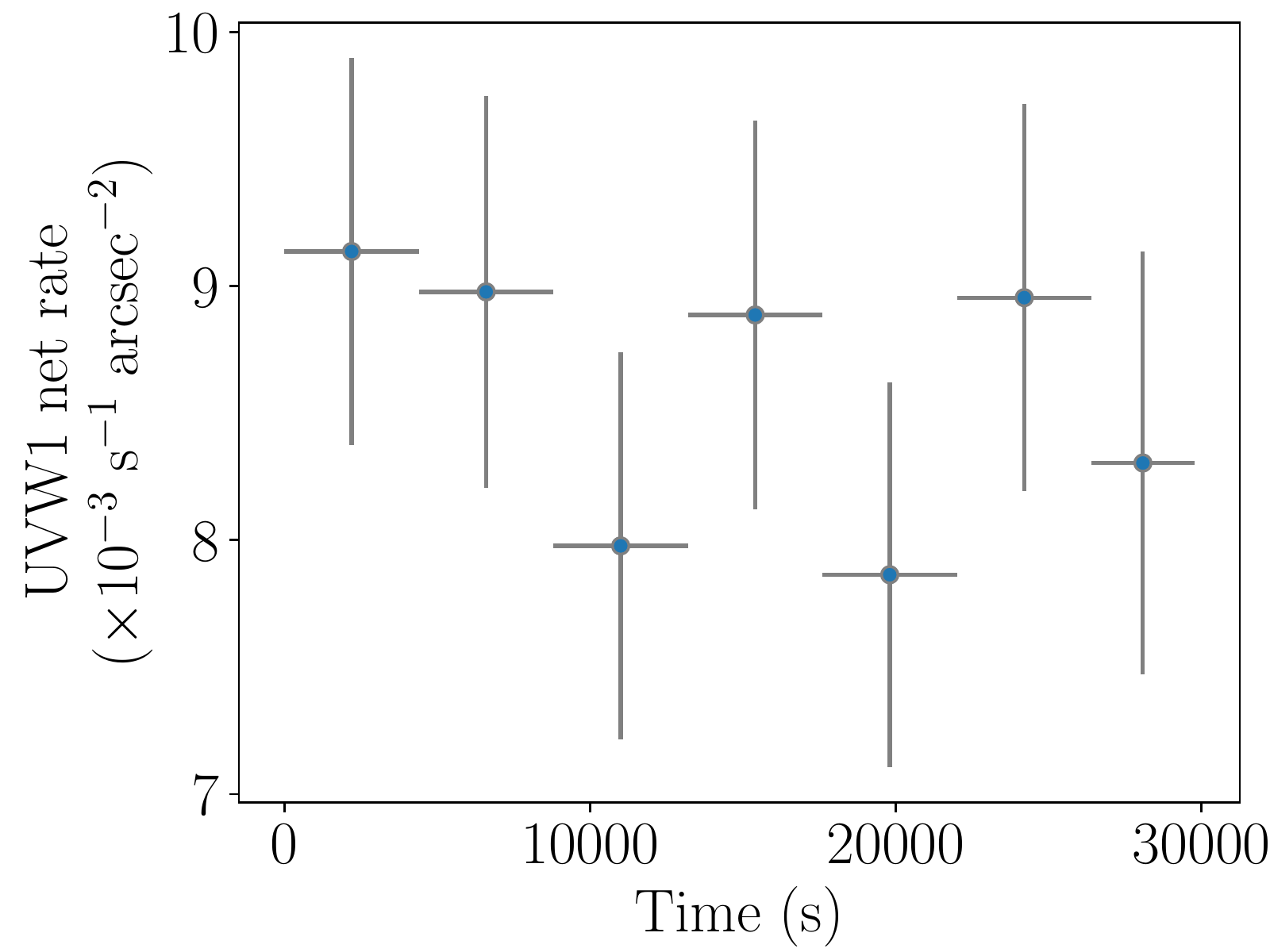}
    \caption{UVW1 background-subtracted light curve of the source as observed by the OM. The seven different exposures performed by the OM are visible. There is no sign of UVW1 variability contemporaneous to the soft X-ray variability.}
    \label{fig:UVW1_light-curve}
\end{figure}

\subsection{Tormund as a QPE source}

\begin{figure}
    \centering
    \includegraphics[width=\columnwidth]{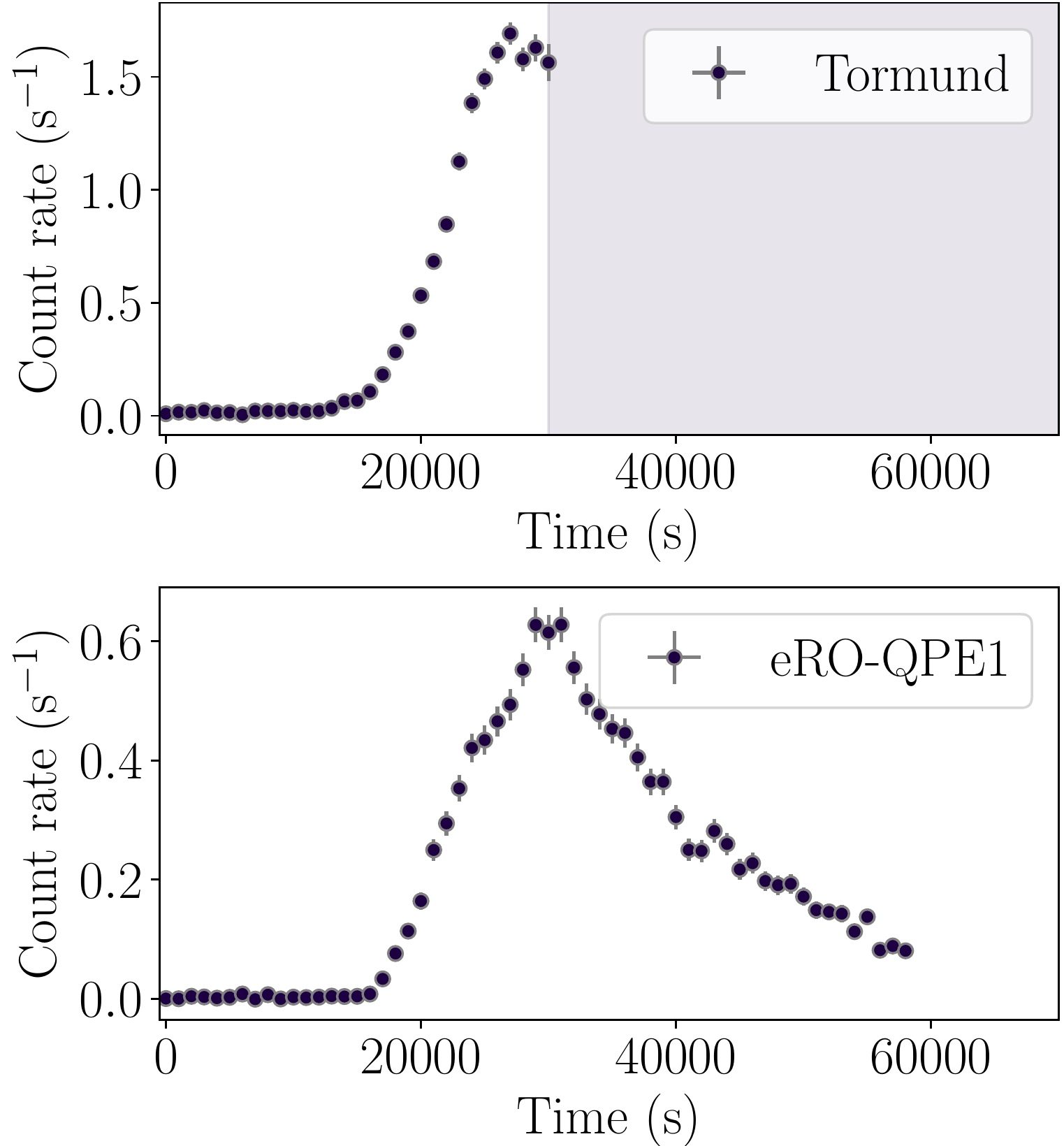}
    \caption{Comparison of EPIC pn 0.3--0.9 keV light curves of Tormund (top panel) and eRO-QPE1 (bottom panel), binned at 1000~s. The first part of the eRO-QPE1 observation was not plotted in order to align the light curves horizontally. The shaded area in the upper panel is after the end of the observation. This shows the similarities between both objects, at least in terms of rising timescales.}
    \label{fig:Tormund_vs_eROQPE}
\end{figure}

Once other interpretations are excluded, one can assess the merits of the QPE interpretation. Since only a partial X-ray burst was detected, Tormund does not qualify immediately as a bona fide QPE source, as this would require the detection of repeating bursts. However, all its properties - luminosity, burst amplitude, thermal spectrum and temperature values, spectral evolution over the burst, and timescales - fit the QPE interpretation, making it a plausible QPE candidate. If multiple peaks had been observed, it would have made Tormund a bona fide QPE source. The fact that only half a probable QPE burst was detected means that Tormund is a candidate QPE source. The short-term spectral evolution of Tormund is remarkably similar to the other known QPEs, with thermal emission increasing steadily from $\sim$50~eV to $\sim$110~eV. The amplitude of the burst, $\sim$125 in count rates, is large but consistent with what is seen in eRO-QPE1 \citep[a factor of 20--300 depending on the bursts;][]{arcodia_x-ray_2021}. The X-ray luminosity of the quiescent state ($\sim10^{42}$~erg~s$^{-1}$) is comparable to that of GSN~069 \citep{miniutti_repeating_2023}, while the luminosity of the peak state ($\sim10^{44}$~erg~s$^{-1}$) is about an order of magnitude larger than those of known QPEs -- the brightest known eruption peak being at the end of the first \textit{XMM-Newton} observation of eRO-QPE1, at $\sim3\times10^{43}$~erg~s$^{-1}$ \citep{arcodia_complex_2022}. Tormund shares two major common features with eRO-QPE1: its large timescales, with a long rise time, and its energy dependence. Indeed, for GSN 069, RX J1301.9+2747, eRO-QPE2, and XMMSL1J024916.604124, the rise time is relatively short (between 2 and 5 ks). In eRO-QPE1, both the rise time and the recurrence time seem to have evolved over the week separating the two \textit{XMM-Newton} observations presented in \citet{arcodia_x-ray_2021}; however, in the second observation, where a single burst was detected, the rise time was around 15 ks in the 0.3--0.9~keV band. This rise time is remarkably similar to that of Tormund. This is shown in Fig.\,\ref{fig:Tormund_vs_eROQPE}, where we have extracted the 0.3--0.9 keV EPIC pn light-curve of eRO-QPE1 from its second \textit{XMM-Newton} observation (ObsID 0861910301) and compared it with that of Tormund. Additionally, the lack of significant variability in the UVW1 light curve (see Fig.\,\ref{fig:UVW1_light-curve}), at least not with the amplitude of the X-ray variability, is also a common feature of QPEs \citep[with the exception of XMMSL1 J024916.604124, where a slight UVW1 dimming was detected at the time of the X-ray bursts;][]{chakraborty_possible_2021}. It is worth noting that a slight optical excess at late times is hinted at in the ZTF r-band light curve in \citet{hammerstein_final_2022}, Fig. 18. This single data point was obtained after host subtraction and time binning over an entire month, which is why it is not present in our light curve in Fig. \ref{fig:all_light-curves}. This point corresponds to about two months after the X-ray brightening of the \textit{XMM-Newton} observation or $\sim$200 days after the optical peak. It is in excess by roughly one order of magnitude of the expected trend, although only at an $\sim2\sigma$ level. If real, this slight variability might be linked to optical reprocessing of the X-ray light. The question of optical reprocessing of X-ray emission from TDEs or QPEs is still open, as none of  the QPEs or the TDEs where a late X-ray counterpart was detected showed any significant optical re-brightening after the X-ray emission \citep[e.g.][]{gezari_x-ray_2017, kajava_rapid_2020,liu_uvoptical_2022}.

If the observed short-term rise in X-ray flux in Tormund was indeed the rise of a QPE, the likelihood of detecting it in a random follow-up would have been higher than for an isolated flare. We showed that a single delayed burst had a low probability of being detected in a non-triggered follow-up. If it was a QPE, however, the fact that they repeat over several months to several years (at least a month for eRO-QPE1, at least 20 years for RX
J1301.9+2747) significantly increases the probability of detecting one. Taking the duty-cycle of eRO-QPE1 \citep[$\sim$40\%,][]{arcodia_x-ray_2021} as a reference due to its similarities with Tormund, we find a probability of about 30\% of detecting at least 20~ks of eruption in a random 30~ks exposure (to be compared with the $\sim0.1\%$ chance of observing the delayed X-ray TDE). 

All of Tormund's observational properties -- luminosity, rise-time, spectrum and spectral evolution, amplitude of variability, and multi-wavelength behaviour -- can be therefore explained with the QPE interpretation, while other interpretations struggle to account for all of the properties. The long rise time for the Tormund burst allowed us to carry out a detailed spectral study. In particular, we can constrain the physical size of the emission region associated with the observed black body. Both of our spectral models lead to an emission radius of $\mathcal{R}_\mathrm{Eruption}=\left(1.30\pm0.05\right) \times10^6$~km. This value can be compared to three different characteristic lengths of our system. The first one is the disc radius in the quiescent state, computed using a $\texttt{bbodyrad}$ component in the first four time windows; it is of roughly the same size as the X-ray eruption region, although less tightly constrained, $\mathcal{R}_\mathrm{Quiescent}=1.04^{+0.62}_{-0.36}\times10^6$~km. The second characteristic length is the size of the central black hole, estimated by taking the mass modelled by \texttt{TDEmass} (resp. \texttt{MOSFiT}) in \citet{hammerstein_final_2022} of 
$M_{\rm BH}^{\tt TDEmass}=6.5^{+2.4}_{-1.7}\times10^6~M_\odot$ (resp. $M_{\rm BH}^{\tt MOSFiT}=8.3^{+0.8}_{-0.7}\times10^7~M_\odot$), yielding a gravitational radius of $\mathcal{R}_{\rm g}^{\tt TDEmass}=9.6^{+3.5}_{-2.5}\times10^6$~km (resp. $\mathcal{R}_{\rm g}^{\tt MOSFiT}=12.2^{+1.2}_{-0.7}\times10^7$~km). While the estimated size of the emission region is here smaller than any of the gravitational radii, which seems unphysical, it is once again important to keep in mind that we might be underestimating the emission radius by up to an order of magnitude \citep{mummery_tidal_2021}. The emission region might indeed be small compared to the gravitational radius if the emission is not directly due to accretion \citep[see e.g.][]{miniutti_repeating_2023,franchini_qpes_2023}. It is also possible that the optical TDE models fail to accurately estimate the black-hole mass \citep[e.g.][]{golightly_diversity_2019}, which is supported by the order-of-magnitude difference in mass estimates between both TDE models. The third characteristic size is the peak black-body radius of the initial optical TDE, computed in \citet{hammerstein_final_2022}, which is $\mathcal{R}_\mathrm{Optical~  TDE}\approx1.2\times10^{10}$~km, about four orders of magnitude larger than the size of the X-ray bright region. This type of behaviour is common in TDEs detected in both visible light and X-rays, with optical emission regions with typical radii three or four orders of magnitude larger than the X-ray emission region (even after accounting for an underestimation by an order of magnitude), with the latter being of comparable size to the gravitational radius or even smaller \citep{gezari_tidal_2021}. One possible explanation for this is that the optical and X-ray emission mechanisms and locations are different, the optical emission being due to shock heating from self-interaction of the debris stream far away from the central black hole, and the X-ray emission arising from delayed accretion once the debris has circularised close to the centre of the system \citep{gezari_tidal_2021}. We can also conclude that the physical extension of the emission region stays roughly constant during the outburst, which is in contrast to what was observed in GSN~069, for instance, with an increase of the black-body radius by a factor of $\sim 2$ over the entire rise and decay \citep{miniutti_repeating_2023}; the increase in GSN~069, however, is most noticeable after the decay phase, which was not observed here. 

\begin{figure}
    \centering
    \includegraphics[width=\columnwidth]{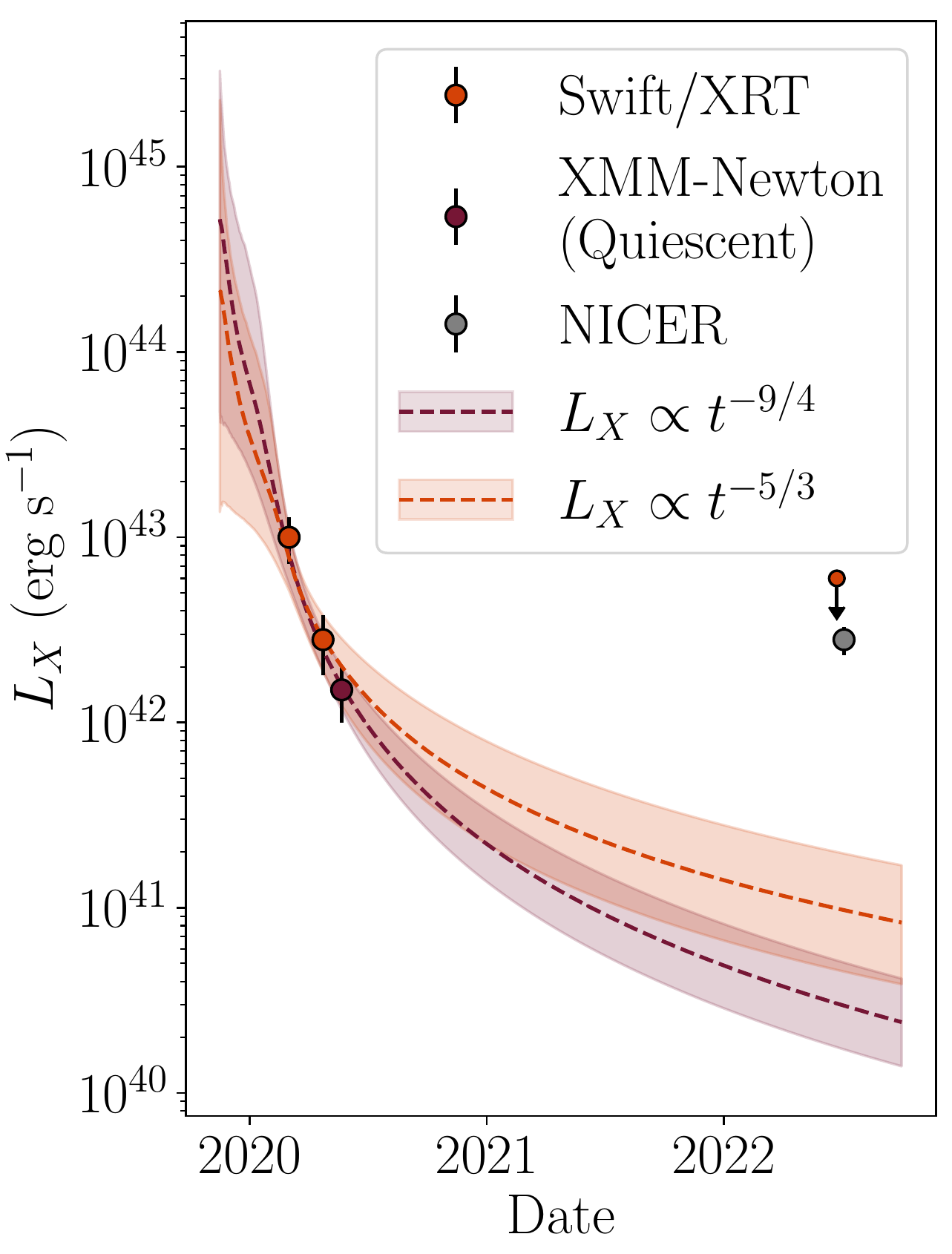}
    \caption{X-ray detections and upper-limits of Tormund, with the expected behaviour of an X-ray TDE following a $L\propto t^{-5/3}$ (respectively $L\propto t^{-9/4}$) evolution in orange (respectively red) based on the first two \textit{Swift} detections and the \textit{XMM-Newton} quiescent state. The dotted lines and shaded areas correspond to the medians, that is the 16$\textsuperscript{th}$ and 84$\textsuperscript{th}$ percentiles of the posterior light curves (see details in Appendix \ref{sec:XTDE}). These contours differ from those of Fig.\,\ref{fig:all_light-curves} in that the \textit{XMM-Newton} quiescent level was also added as data for the light-curve fitting, allowing for better constrained envelopes. We note that the outburst during the second half of the \textit{XMM-Newton} observation is not shown here, as we are only interested in the evolution of the quiescent state, which is expected to behave in a TDE-like fashion. The \textit{NICER} detection is in excess of the expected behaviour of the quiescent level for both power-law decay indices, hinting at a possible re-brightening.}
    \label{fig:XrayRebrightening}
\end{figure}

The late-time \textit{NICER} detection of a soft emission provides us with additional information. No clear sign of variability was detected during the \textit{NICER} exposure between the approximately two-day long snapshots, although the relatively low signal-to-noise ratio prevented a finely time-resolved approach. The combined emission in the \textit{NICER} observations, at $2.8^{+0.48}_{-0.48}\times10^{42}$ erg~s$^{-1}$, is brighter than expected. Indeed, \citet{miniutti_repeating_2023} showed that the quiescent state of GSN~069 roughly followed at first the $L\propto t^{-9/4}$ expected from a partial TDE \citep{coughlin_partial_2019} and then underwent a re-brightening. We simulated the same behaviour for Tormund, with possible X-ray TDE light curves following either $L\propto t^{-9/4}$ or $L\propto t^{-5/3}$. We used the first two \textit{Swift}/XRT detections and the \textit{XMM-Newton} quiescent state as data, and the method and priors detailed in Appendix \ref{sec:XTDE}. The X-ray detections of the source and the simulated light curves can be found in Fig.\,\ref{fig:XrayRebrightening}. The \textit{NICER} detection is in excess by a factor of $28^{+43}_{-17}$ of the $L\propto t^{-5/3}$ light curve and by a factor of $95^{+97}_{-49}$ of the $L\propto t^{-9/4}$ light curve. There are several possible interpretations for this excess. If the QPEs are still active, the \textit{NICER} detection corresponds to an average of peaks and quiescent state (the low signal preventing us from clearly observing the variability), which could lead to this excess. If the QPEs are no longer active, the \textit{NICER} detection corresponds to a quiescent state, which would then not have followed the expected TDE-like behaviour observed in GSN~069. In particular, this excess could be explained by a re-brightening, similar to what was witnessed in GSN~069, which could have happened anytime between the \textit{XMM-Newton} and the \textit{NICER} detections; the lack of continuous X-ray coverage prevents us from making any strong conclusions. Such a re-brightening would need to have had a much larger amplitude than for GSN~069 (which re-brightened by a factor of $\sim$2), which would be consistent with the fact that the QPE amplitudes are also larger in Tormund.

Even if the observed \textit{XMM-Newton} burst was not the start of a QPE but a single isolated TDE flare (which we argue is unlikely at the start of this section), the late-time optical spectrum excluded the \textit{NICER} detection from being due to a quiescent AGN emission. This tells us that this X-ray TDE is still active more than 900 days after the optical TDE, and about 750 days after the observed X-ray burst; this large duration is to be compared to the typical $\sim$100 day optical duration of X-ray-bright TDEs \citep{hammerstein_final_2022}. 

\subsection{What this tells us about QPEs in general}

We show that the best interpretation for the rapid increase of soft X-ray flux witnessed in Tormund is that it was the rising phase of a QPE. It thus joins the group of strong QPE candidates, along with XMMSL1 J024916.6-041244; for both of them, only the low number of observed eruptions prevents us from concluding their nature with absolute certainty. This new addition has two major effects: extending the parameter space of QPEs and strengthening the observational link between QPEs and TDEs.

In terms of 0.2--2~keV luminosity, Tormund is the brightest of all QPEs for peak luminosity ($\sim 1.2\times10^{44}$~erg~s$^{-1}$), with quiescent luminosity comparable to the other QPEs ($\sim10^{42}$~erg~s$^{-1}$). The central black-hole is also the most massive of the sample, with log($M_{\rm BH}/\rm M_\odot)=6.8\pm0.13$ \citep{hammerstein_final_2022}, compared to eRO-QPE1 with log($M_{\rm BH}/\rm M_\odot)=5.78\pm0.55,$ for instance \citep{wevers_host_2022}. This hints at a link between the black-hole mass, the QPE luminosity, and the typical rise time (as neither decay nor recurrence time were observed here). These parameters could be tied, for instance, through the size of the eruption region, which could be compared between the various QPEs, with $\mathcal{R}_{\mathrm{Eruption}}\sim10^{6}$~km for both Tormund and eRO-QPE1, compared to $\sim2\times10^{5}$~km for GSN~069 \citep{arcodia_complex_2022,miniutti_repeating_2023}. Another possible explanation for the high luminosity could be the short time since the initial TDE, at least compared to GSN~069 and XMMSL1 J024916.6-041244, which could suggest that there is still a large quantity of matter available to interact with (especially in the disc-collision model).

The direct link of Tormund as a QPE with a TDE strengthens the observational correlation between these two phenomena, which are most likely physically linked. We can try to estimate the probability of randomly observing such a sample of events, exhibiting both TDE followed by QPEs. From the $\sim$600\,000 sources in the 4XMM-DR11 catalogue, about 75\% are nuclear sources \citep{tranin_probabilistic_2022}. With an estimated TDE rate of $\sim6\times10^{-5}$~yr$^{-1}$~galaxy$^{-1}$ \citep{van_velzen_optical-ultraviolet_2020}, over the 20 year coverage of 4XMM-DR11, this leads to an expected number of observed TDEs of about 550 in the 4XMM-DR11 catalogue. This means that a random nuclear source from the 4XMM-DR11 catalogue has a 550/450 000 = 0.12\% of being a TDE. With Tormund, the sample of QPEs (bona fide or candidates) is increased to six, three of them being linked to past TDEs. Assuming that these two physical phenomena are completely independent, the probability of witnessing such a correlation purely randomly is $(0.12\%)^{2}=1.4\times10^{-6}$ before the discovery of Tormund, and $(0.12\%)^{3}=1.7\times10^{-9}$ when taking into account this new QPE source. This extremely low probability of a random correlation strongly favours physical models where the QPE phenomenon is linked with the TDE one \citep[e.g.][]{king_gsn_2020,xian_x-ray_2021,wang_model_2022}. 
 
The various models for QPEs rely on different formation channels and emission processes (see Sect.\,\ref{sec:Intro}), so their estimates for the formation time and the lifetime of QPEs might differ, and constraining those might help exclude or strengthen some models. Until now, the focus has been mostly put on the lifetime of QPEs, as a regular monitoring of active QPEs easily allows us to constrain it. The lifetime in the disc collision model is limited by the existence of the underlying TDE-created disc, typically lasting a few months to a few years \citep{xian_x-ray_2021}. The model of QPEs as tidal stripping of a white dwarf requires the survival of the orbiting white dwarf, giving a typical lifetime of $\sim$2 years once a luminosity of $10^{42}$~erg~s$^{-1}$ is reached \citep{wang_model_2022}. For the model of two coplanar counter-orbiting EMRIs \citep{metzger_interacting_2022}, QPEs stop once Lense-Thirring nodal precession leads to a misalignment of the orbits; after a few months -- once they are aligned again -- QPEs might start again. The observations seem to be consistent with the models in terms of lifetime. \citet{miniutti_repeating_2023} reported on the disappearance of QPEs in GSN~069 after $\sim 1$ year of activity, associated with a significant re-brightening of the quiescent state interpreted as a second partial TDE and predicted the future reappearance of QPEs. For RX J1301.9+2747, the QPEs have been active for at least 20 years \citep{giustini_x-ray_2020}, which would require us to adjust the models for them to be active for so long, for instance by changing the donor to a post-AGB star instead of a white dwarf in the tidal stripping model \citep{zhao_quasi-periodic_2022}. 
For Tormund, the recent \textit{NICER} detection allows us to confirm that the source is still active with soft emission at a level of about $2\times10^{42}$~erg~s$^{-1}$, but the necessary stacking of the snapshots prevents any conclusion on the current presence of QPEs. 

Regarding the formation time, Tormund is the first QPE-candidate for which the associated TDE was directly detected, and not simply deduced from variability from an archival quiescent flux or optical spectral features (as was the case for GSN 069 and XMMSL1 J024916.6-041244). This means that this is the first QPE-candidate with strong constraints on the formation time of QPEs after the TDE, constrained to below six months in the case of Tormund, compared to at least four years for GSN~069 and at least two years for XMMSL1 J024916.6-041244 \citep{miniutti_nine-hour_2019,chakraborty_possible_2021}. For the model of tidal stripping of a white dwarf, the QPEs appear once the loss of orbital energy through gravitational waves brings the white dwarf close to the central black hole to trigger Roche lobe overflows; this is estimated to take up to a few years \citep{wang_model_2022}, which is different to what was observed in Tormund. The model of coplanar counter-orbiting EMRIs \citep{metzger_interacting_2022} requires an almost total circularisation of the remnant from the initial TDE through the emission of gravitational waves as well, on a typical timescale of several years, which seems inconsistent with Tormund. For the disc collision model, the formation time of QPEs after the TDE will depend on the time taken to circularise the debris on a misaligned orbit with respect to the remnant. If the accretion disc is formed from the disrupted envelope of the star, it is initially coplanar with the remnant. For QPEs to appear due to collisions, the orbital planes of the disc and the remnant need to evolve differently, most likely through frame dragging around the rotating central black hole. The typical Lense-Thirring nodal precession is $\sim$0.01$\pi$ per orbit \citep{hayasaki_circularization_2016}, which is fast enough to change the orbital planes within the six-month constraint. The repeated interactions between the disc and the remnant would change the orbital parameters of the latter, and thus of the recurrence times. A more precise estimation of these phenomena would require us to account for the interactions within the disc, as different radii experience different levels of frame dragging, and to account for the presence of the remnant that would perturb the debris orbits \citep{wang_partial_2021}. 

The constraints on formation time add a new parameter to the increasing list of observational features of QPEs that models need to account for. The current major observed properties of QPEs, including both confirmed QPE sources and QPE candidates, are as follows: their X-ray 0.2--2 keV luminosities are within $10^{40}$--$10^{42}$~erg~s$^{-1}$ for the quiescent state and reach $10^{42}$--$10^{44}$~erg~s$^{-1}$ at the peak. They are characterised by a soft X-ray spectrum, consistent with a black body heating up from $\sim$50~eV to $\sim$100~eV, with no other significant multi-wavelength counterpart. Their bursts last from $\sim$5~ks to $\sim$35~ks, with a pulse profile that can be either rather symmetrical (e.g. GSN~069) or asymmetrical (eRO-QPE1), with the caveat that asymmetry is statistically harder to confirm for short timescales. The recurrence time between bursts evolves over time, sometimes being smaller than the burst duration, leading to overlapping peaks (a single QPE source can change pulse type in less than a week; eRO-QPE1). The bursts show an apparent pattern of smaller and larger peaks, and shorter and longer recurrence times alternating in GSN~069 and eRO-QPE2, 
or more complex and irregular behaviour in eRO-QPE1 and RX J1301.9+2747. Their long-term behaviour is characterized by a lifetime ranging from below two years (GSN~069) to over 20 years (RX J1301.9+2747), and a disappearance of the QPEs sometimes associated with a re-brightening of the quiescent state (GSN~069). Quasi-periodic oscillations in the quiescent state in GSN~069, at the same period as the QPEs, have been observed. In terms of host properties, they have all been detected around low-mass SMBHs, with masses in the range of $10^5$ -- $10^7~M_\odot$. Finally, there is a strong observational correlation with TDEs, with a formation time after the TDE that can be as short as a few months in the case of Tormund.
All the currently proposed models struggle with at least some observational properties, among which are the changing burst profile and the recent detection of QPOs in the quiescent state of GSN~069 \citep{miniutti_repeating_2023}.

\section{Conclusions}
In this work, we present a detailed study of the short-term X-ray variability witnessed in an \textit{XMM-Newton} follow-up of the optically detected TDE AT2019vcb, nicknamed Tormund. Before the \textit{XMM-Newton} outburst, two prior detections of very soft variable X-ray emission by \textit{Swift}/XRT were consistent with an X-ray-bright TDE. As we only detected the rise phase of one isolated QPE-like feature, Tormund cannot qualify as a bona fide QPE source. However, the properties of Tormund's variability event reported here are strikingly similar to those of other QPE sources, and all other interpretations struggle in accounting for all observational features. The similarities of Tormund with known QPEs, especially with eRO-QPE1, in terms of spectral-timing properties (a black body heating up from $53.5^{+9.2}_{-7.7}$~eV to $113.8^{+2.9}_{-2.7}$~eV) and luminosity (ranging from $L^{\mathrm{Quiescent}}_{\rm 0.2-2keV}=3.2^{+1.6}_{-1.0}\times10^{42}$\,erg\,s$^{-1}$ to $L^\mathrm{Peak}_{\rm 0.2-2keV}=1.19^{+0.05}_{-0.05}\times10^{44}$\,erg\,s$^{-1}$) are in favour of the interpretation that Tormund hosted QPEs, despite only the rising phase of a single eruption having been detected at the time. This lead us to the conclusion that Tormund deserves to be given the status of candidate QPE source.

This interpretation would allow several constraints to be put on our current understanding of QPEs:
\begin{itemize}
    \item This detection increases the fraction of TDE-linked QPEs from two out of five to three out of six. Considering the rarity of TDEs among X-ray sources, this strengthens the case for a strong link between QPEs and TDEs. The quiescent state, right before the sudden X-ray increase, is consistent with the decay phase of the TDE. It is worth pointing out that, for the three remaining QPEs with no clear link to a TDE, a prior TDE is not excluded but simply not detected;
    \item This is the first QPE candidate that was detected after a clearly observed optical TDE with a well-determined optical peak. This gives us a stronger upper limit on the formation time of QPEs, which must be below six months in the case of Tormund. The spectral classification of the optical TDE, H+He, is consistent with an evolved star, which would be more likely to lead to a partial TDE. Repeated TDEs, possibly associated with the partial disruption of the envelope of an evolved star, are also inferred from the long-term X-ray evolution of GNS~069 \citep{miniutti_repeating_2023}.
    \item The evolution of the eruption flux (total flux minus quiescent flux) is consistent with heating a constant-sized region, with a typical physical scale of $R_{\text{Eruption}}=\left(1.3\pm0.05\right)\times10^{6}$~km, which is about five times smaller than the gravitational radius of the central black hole and four orders of magnitude smaller than the initial optical TDE, meaning that the emission region of QPEs is extremely limited in size. The eruption region appears comparable in size to the quiescent emission, although the latter might be severely underestimated because of various scattering effects \citep{mummery_tidal_2021}.
    \item Among the sample of known QPEs (bona fide and candidates), Tormund has the largest rising timescale ($\sim15$ ks, similar to eRO-QPE1), and the most massive central black hole mass ($\sim6.5\pm1.5\times10^6~ M_\odot$), perhaps hinting at a correlation between those properties; it would also replace eRO-QPE1 as the brightest and most distant QPE to date. The lack of decay phase or further bursts prevents conclusions on the other typical timescales of QPEs. 
    \item The late \textit{NICER} detection of soft emission indicates that the source is still active $\sim$2 years after the first outburst, although the signal is too weak to provide any conclusion on the presence of QPEs. Thanks to the late-time optical spectrum, we can exclude any possible AGN activity. The \textit{NICER} detection therefore leads to the conclusion that the TDE-linked X-ray emission has been lasting for over 900 days after the optical peak.
\end{itemize}

One of the main obstacles to improving our current understanding of QPEs is the very low available sample of candidates. Tormund provides us with a new candidate, which also broadens the parameter space in terms of luminosity, timescales, and central black-hole mass; it is also the first QPE candidate for which a strong formation time constraint can be estimated. This will allow ulterior models and simulations to have tighter constraints and hopefully help us understand the precise emission mechanisms behind these phenomena. We will continue our long-duration monitoring of this source with \textit{Swift} in order to confirm the disappearance -- or reappearance -- of QPEs in Tormund. Finally, a possible avenue to find additional QPE candidates is to use more complete galaxy catalogues than the one used in this study, as it was shown that one of the known QPEs (XMMSL1
J024916.6-04124) was not in this catalogue; in particular, we intend to make use of the Gaia DR3 catalogue \citep{carnerero_gaia_2022}.




\begin{acknowledgements}

Softwares: \texttt{numpy} \citep{harris_array_2020}, \texttt{matplotlib} \citep{hunter_matplotlib_2007}, \texttt{astropy} 
\citep{astropy_collaboration_astropy_2013,astropy_collaboration_astropy_2018,astropy_collaboration_astropy_2022}, \texttt{PYRAF}\footnote{\url{https://iraf-community.github.io/pyraf.html}}, \texttt{PyMC} \citep{salvatier_probabilistic_2016}, \texttt{CMasher} \citep{van_der_velden_cmasher_2020}, \texttt{scipy} \citep{virtanen_scipy_2020}, \texttt{Xspec} \citep{arnaud_xspec_1996}, \texttt{SAS}, \texttt{NICERDAS}.
The authors thank the anonymous referee for useful comments that helped improve the quality of this paper. Some of this work was done as part of the XMM2ATHENA project. This project has received funding from the European Union's Horizon 2020 research and innovation programme under grant agreement n°101004168, the XMM2ATHENA project. EQ, NAW, SG, EK, NC and RAm acknowledge the CNES who also supported this work. Some of the results were based on observations obtained with the Samuel Oschin Telescope 48-inch and the 60-inch Telescope at the Palomar Observatory as part of the Zwicky Transient Facility project. ZTF is supported by the National Science Foundation under Grant No. AST-2034437 and a collaboration including Caltech, IPAC, the Weizmann Institute for Science, the Oskar Klein Center at Stockholm University, the University of Maryland, Deutsches Elektronen-Synchrotron and Humboldt University, the TANGO
Consortium of Taiwan, the University of Wisconsin at Milwaukee, Trinity College Dublin, Lawrence Livermore National Laboratories, and IN2P3, France. Operations are conducted by COO, IPAC, and UW. MG is supported by the ``Programa de Atracci\'on de Talento'' of the Comunidad de Madrid, grant number 2018-T1/TIC-11733. R.Ar acknowledges support by NASA through the NASA Einstein Fellowship grant No HF2-51499 awarded by the Space Telescope Science Institute, which is operated by the Association of Universities for Research in Astronomy, Inc., for NASA, under contract NAS5-26555. This work is based in part on observations made at Observatoire de Haute Provence (CNRS), France.

\end{acknowledgements}

\bibliographystyle{aa}
\bibliography{references}

%
%
\begin{appendix}
\section{Computation of the X-ray TDE light curves}
\label{sec:XTDE}
To compute the possible light curves of the X-ray TDE prior to the \textit{XMM-Newton} burst, we compared the first two  \textit{Swift}/XRT observations to the standard $L_{X}~\propto~t^{-5/3}$ evolution of TDE luminosity over time \citep[e.g.][]{gezari_tidal_2021}. The precise model we used was $L_{X}~=L_{\mathrm{Peak,X}} \times ((t-t_{\mathrm{Peak,X}}+t_{0})/t_{0})^{-5/3}$, with $L_{\mathrm{Peak,X}}$, $t_{0}$ and $t_{\mathrm{Peak,X}}$ free parameters \citep{van_velzen_seventeen_2021}. No quiescent constant component was added, as there is no indication in the optical spectrum of the presence of an AGN. The parameters and their flat priors are summarised in Table\,\ref{tab:Xtde_model}. To estimate these parameters and the possible X-ray TDE light curves, we used the \texttt{PyMC} framework \citep{salvatier_probabilistic_2016} with a Gaussian likelihood function and the \texttt{NUTS} sampler. We used 25 walkers on 8~000 steps, discarding the first 2~000. The median 16$\textsuperscript{th}$ and 84$\textsuperscript{th}$ percentiles of the associated posterior light curves were computed for each time step, shown by the orange dotted line and shaded area in Fig.\,\ref{fig:all_light-curves}, respectively. While the median line does not correspond in itself to a TDE light curve, it provided us with a rough estimate of the X-ray behaviour of the source prior to the \textit{XMM-Newton} burst. Changing the power-law index to $-9/4$ instead of $-5/3$, which would correspond to a partial TDE \citep[e.g.][]{coughlin_partial_2019}, does not affect the shape of the envelope at the time of the \textit{XMM-Newton} observation. 

This method proved that the \textit{XMM-Newton} quiescent state was consistent with the median of the posterior light curves, that is consistent with the same TDE decay that was seen by \textit{Swift}. Adding this third data point to the fitting procedure, we repeated this approach, this time with both power-law indices. This lead to tighter envelopes, which are shown in Fig.\,\ref{fig:XrayRebrightening}.

\begin{table}[h]
\renewcommand{\arraystretch}{1.3}
\centering
\begin{tabular}{cccc}
\hline \hline
Parameter & Description & Prior \\
\hline
log $L_{\mathrm{Peak}}$ & Peak X-ray luminosity & [42,46] erg~s$^{-1}$ \\
log $t_{0}$ & Power-law normalisation  & [-4,0.5] years \\
$t_{\mathrm{Peak,X}}$ & Time of peak & [0,108]  \\
\hline
\end{tabular}
\caption{Free parameters used for modelling the X-ray TDE light curve. The time of peak $t_{\mathrm{Peak,X}}$ is expressed with respect to the date of the optical alert (November 15$\textsuperscript{th}$, 2019) -- its upper limit of 108 days corresponds to the first \textit{Swift}/XRT detection, in which the source already behaved like a TDE (March 1$\textsuperscript{st}$, 2020). For the prior of the peak luminosity, the softness of the X-ray emission allows to rule out a jetted TDE \citep[e.g.][]{pasham_birth_2023}, which would be necessary to reach observed luminosities above $10^{46}$~erg~s$^{-1}$.}
\label{tab:Xtde_model}
\end{table}

\begin{figure*}
    \centering
    \includegraphics[width=\textwidth]{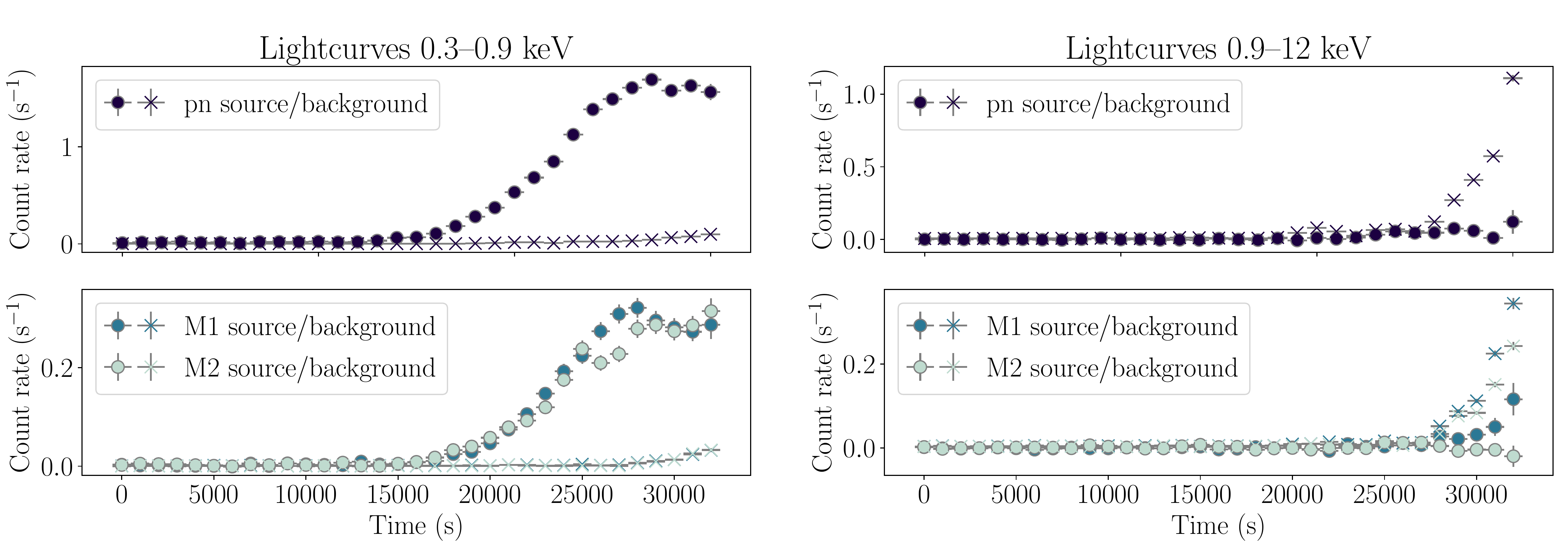}
    \caption{\textit{XMM-Newton} source and background light curves of Tormund, observation 0871190301, for the three EPIC instruments and in two energy bands. The left panels correspond to energies below 0.9 keV, and the right panels to energies above 0.9 keV. The top panels show the EPIC pn light curves and the bottom panels show both the EPIC MOS1 and MOS2 light curves. The background-subtracted light curves are shown with filled circles, while the backgrounds are shown with crosses; backgrounds are scaled to the area of the source extraction region. This figure shows that a background flare is present at the end of the observation, visible in both low and especially high energies, but there is definitely an intrinsic source variability in the soft energy band, with a high signal-to-noise ratio even during the flare. The extreme softness of the source allows us to reasonably keep time windows in which the high-energy flaring background is above the recommended threshold, under the condition that we discard any data above 0.9 keV. Data below 0.3 keV are discarded to avoid calibration issues between EPIC instruments.}
    \label{fig:EPIClight-curves}
\end{figure*}

\begin{table*}
\centering
\begin{tabular}{c|cccccc}
\hline \hline
Energy band & Quiescent rate ($\times 10^{-2}$ s$^{-1}$) & Start time (h) & Rise time (h) & Peak time (h) & Peak rate ($\times 10^{-2}$ s$^{-1}$) & $\chi^{2}$/DoF \\ \hline \hline
0.30 -- 0.45 keV & $0.79\pm0.13$& $3.9\pm0.05$& $3.76\pm0.14$& $7.66\pm0.19$& $96.4\pm2.2$ & 70.3 / 58 \\
0.45 -- 0.60 keV & $0.36\pm0.08$& $4.48\pm0.04$& $2.96\pm0.13$& $7.44\pm0.17$& $67.7\pm1.6$ & 57.8 / 54 \\
0.60 -- 0.75 keV & $0.21\pm0.09$& $4.95\pm0.05$& $2.24\pm0.15$& $7.19\pm0.2$& $36.8\pm1.1$ & 44.5 / 40 \\
0.75 -- 0.90 keV & $0.23\pm0.05$& $5.15\pm0.06$& $1.92\pm0.14$& $7.07\pm0.2$& $18.6\pm0.5$ & 20.9 / 39 \\ \hline
0.30 -- 0.90 keV & $1.68\pm0.16$& $4.3\pm0.03$& $3.27\pm0.08$& $7.56\pm0.11$& $221.4\pm3.1$ & 61.9 / 58 \\ \hline
\end{tabular}
\caption{Results of the fitting of the light curves of the \textit{XMM-Newton} burst in different energy bands, with their $1\sigma$ uncertainties. The burst model used was a Gaussian rise between two plateau phases. The start and peak times are expressed with respect to the start of the observation.}
\label{tab:GaussianFitlight-curves}
\end{table*}

\begin{table*}
\centering
\renewcommand{\arraystretch}{1.5}
\begin{tabular}{ccccc}
\hline \hline
Time window & kT & Norm  & $L_{\text{0.2-2 keV}}$ & Radius  \\
&(eV)&$\left( \times10^{-6}~L^{\mathrm{bol}}_{39}D^{2}_{10}\right)$&$t( \times10^{42}$~erg~s$^{-1})$ &($\times10^{6}$ km)\\\hline \hline
0 (1-4 combined) &70.1$^{+8.9}_{-8.11}$&1.81$^{+1.0}_{-0.6}$&3.2$^{+1.6}_{-1.0}$&1.04$^{+0.62}_{-0.36}$\\ \hline
1&100.8$^{+37}_{-23}$&0.6$^{+0.57}_{-0.25}$&1.1$^{+1.0}_{-0.5}$&0.29$^{+0.39}_{-0.17}$\\
2&63.8$^{+23}_{-17}$&2.3$^{+7.5}_{-1.4}$&4.0$^{+11.4}_{-2.4}$&1.42$^{+2.66}_{-0.94}$\\
3&78.4$^{+23}_{-17}$&1.2$^{+1.6}_{-0.6}$&2.2$^{+2.6}_{-1.0}$&0.69$^{+1.02}_{-0.38}$\\
4&56.1$^{+16}_{-14}$&5.0$^{+18.5}_{-3.2}$&8.4$^{+26.8}_{-5.2}$&2.71$^{+7.46}_{-1.71}$\\
5&53.5$^{+9.2}_{-7.7}$&11.9$^{+15.1}_{-5.9}$&19.7$^{+22.3}_{-9.5}$&4.58$^{+4.76}_{-392.2}$\\
6&68.5$^{+5.1}_{-4.8}$&15.1$^{+4.6}_{-3.3}$&26.6$^{+7.5}_{-5.5}$&3.16$^{+1.0}_{-0.73}$\\
7&85.9$^{+3.7}_{-3.5}$&26.5$^{+3.4}_{-2.9}$&48.3$^{+6.0}_{-5.1}$&2.67$^{+0.41}_{-0.35}$\\
8&106.5$^{+3.0}_{-2.9}$&38.7$^{+2.2}_{-2.1}$&71.7$^{+4.0}_{-3.7}$&2.09$^{+0.17}_{-0.16}$\\
9&113.1$^{+2.6}_{-2.5}$&58.5$^{+2.4}_{-2.3}$&108.6$^{+4.5}_{-4.2}$&2.28$^{+0.14}_{-0.14}$\\
10&113.8$^{+2.9}_{-2.7}$&64.1$^{+2.8}_{-2.7}$&118.9$^{+5.3}_{-4.9}$&2.35$^{+0.16}_{-0.15}$ \\ \hline
\end{tabular}
\caption{Spectral best-fit parameters of the ten time windows of the \textit{XMM-Newton} observation,  using \texttt{tbabs$\times$zbbody}, with their 90\% confidence intervals. The time window 0 results from the merging of the time windows 1--4 and corresponds to the quiescent state. The bolometric black-body luminosity is computed by using the normalisation. The 0.2--2~keV luminosity is then computed by using the temperature obtained from the fitted data in the 0.3--0.9~keV band. The radius is computed by replacing \texttt{zbbody} with \texttt{zashift}$\times$\texttt{bbodyrad}, which allows us to retrieve the emitting area in the normalisation factor. The spectral fitting of the time windows 1--10 resulted in a fit statistic of $C_{Stat}=341.48$ with 341 degrees of freedom. The combined quiescent state fit yields a fit statistic of $C_{Stat}=34.64$ with 33 degrees of freedom. For slices 0, 5, and 6, the low signal prevents us from using $C_{Stat}$ as a direct estimate of the goodness of fit. The Monte Carlo simulations of the best-fit spectra we performed confirmed the quality of the fit. We found percentages of the worst realisation of the fits of 12\% for the quiescent state, and of 68\% and 4\% for the first two eruption slices, respectively, the latter being marginally acceptable.}
\label{tab:fitbbody}
\end{table*}

\begin{table*}
\centering
\renewcommand{\arraystretch}{1.5}
\begin{tabular}{ccccccc}
\hline \hline
Time window & kT$_{\text{Diskbb}}$(eV) &Norm$_{\text{Diskbb}}$&kT$_{\text{Bbody}}$ & Norm$_{\text{Bbody}}$  & $L_{\text{0.2-2 keV}}$  & Radius \\
& (eV) & $\left( (R_{in}/D_{10})^{2} \right) $&(eV) &$\left( \times10^{-6}~L^{\mathrm{bol}}_{39}D^{2}_{10}\right)$ & ($\times10^{42}$~erg~s$^{-1}$) & ($\times 10^6$~km)\\\hline \hline
0 (1-4 combined) & $84^{+12}_{-11}$ & $299^{+580}_{-190}$ & N/A &N/A &N/A & N/A\\ \hline
1&$86^{+21}_{-17}$&$206^{+637}_{-122}$& & & & \\
2&"&"& &  & & \\
3&"&"& &  &  &\\
4&"&"& & && \\
5&"&"&45.1$^{+16.2}_{-10.1}$&17.1$^{+83.6}_{-13.1}$&26.4$^{+105.2}_{-19.6}$&8.5$^{+29.8}_{-5.9}$\\
6&"&"&68.0$^{+5.8}_{-5.5}$&13.8$^{+5.1}_{-3.8}$&24.3$^{+8.3}_{-6.4}$&3.1$^{+1.2}_{-0.8}$\\
7&"&"&86.3$^{+3.9}_{-3.7}$&25.3$^{+3.4}_{-2.8}$&46.1$^{+6.0}_{-4.9}$&2.6$^{+0.4}_{-0.4}$\\
8&"&"&107.1$^{+3.1}_{-3.0}$&37.8$^{+2.2}_{-2.0}$&70.1$^{+4.0}_{-3.6}$&2.1$^{+0.2}_{-0.2}$\\
9&"&"&113.5$^{+2.6}_{-2.5}$&57.7$^{+2.4}_{-2.2}$&107.1$^{+4.5}_{-4.1}$&2.3$^{+0.1}_{-0.1}$\\
10&"&"&114.2$^{+2.9}_{-2.8}$&63.3$^{+2.8}_{-2.6}$&117.5$^{+5.2}_{-4.8}$&2.3$^{+0.2}_{-0.2}$ \\ \hline
\end{tabular}
\caption{Spectral fitting parameters of the ten time windows of the \textit{XMM-Newton} observation, using \texttt{tbabs$\times$zashift$\times$(bbody+diskbb)}, with their 90\% confidence intervals. The Time window 0 results from the merging of the time windows 1-4, and corresponds to the quiescent state; we removed the \texttt{bbody} parameter from this specific time window only, to better estimate the quiescent \texttt{diskbb}. For the other time windows, the \texttt{diskbb} parameters are tiedn but the \texttt{bbody} is independent. The luminosity is for the \texttt{bbody} only, and it is extrapolated to the 0.2-2 keV band from the fitted data in the 0.3-0.9 keV band. It was not computed in the first four time windows, where the \texttt{bbody} was not constrained. The radius is computed by replacing \texttt{bbody} with \texttt{bbodyrad}, which allows us to retrieve the emitting area from the normalisation factor. The spectral fitting of the time windows 1--10 yields a fit statistic of $C_{Stat}=339.4$ with 339 degrees of freedom. The combined quiescent state fit resulted in a fit statistic of $C_{Stat}=32.7$ with 33 degrees of freedom.}
\label{tab:fitdiskbb}
\end{table*}

\begin{table*}
\centering
\renewcommand{\arraystretch}{1.3}
\begin{tabular}{c|cc|cc}
\hline \hline
Model &$ \alpha$ & $\chi^{2}$~/~D.o.F & $\chi^{2}$~/~D.o.F for $\alpha=4$ & Inferred Radius \\ \hline \hline
\texttt{bbody} - with quiescent& $4.18\pm0.95$ & 13.04~/~5 & 13.14~/~6 & N/A \\
\texttt{bbody} - without quiescent& $3.28\pm0.63$ & 5.98~/~4 & 7.50~/~5 & $(1.30\pm0.05) \times10^6$~km\\
\texttt{diskbb}+\texttt{bbody} & $3.55\pm0.69$ & 5.29~/~4 & 5.78~/~5 & $(1.27\pm0.04) \times10^6$~km \\ \hline
\end{tabular}
\caption{Fitting parameters of luminosity versus temperature behaviour of the source during the \textit{XMM-Newton} observation, with both the \texttt{bbody} model and the \texttt{diskb+bbody} model. For the \texttt{bbody} model, the quiescent state is clearly an under-luminous outlier (see Fig.\,\ref{fig:EruptionFluxVersusT}), so we performed the fitting both including and excluding the quiescent state. Two fits were performed for each model: one to a power-law behaviour $L\propto T^{\alpha}$ with $\alpha$ free and one when fixing $\alpha=4$ (constant-sized black body) and fitting only the proportionality factor, which is linked to the emission region radius.}
\label{tab:LversusT}
\end{table*}

\begin{figure}
    \centering
    \includegraphics[width=\columnwidth]{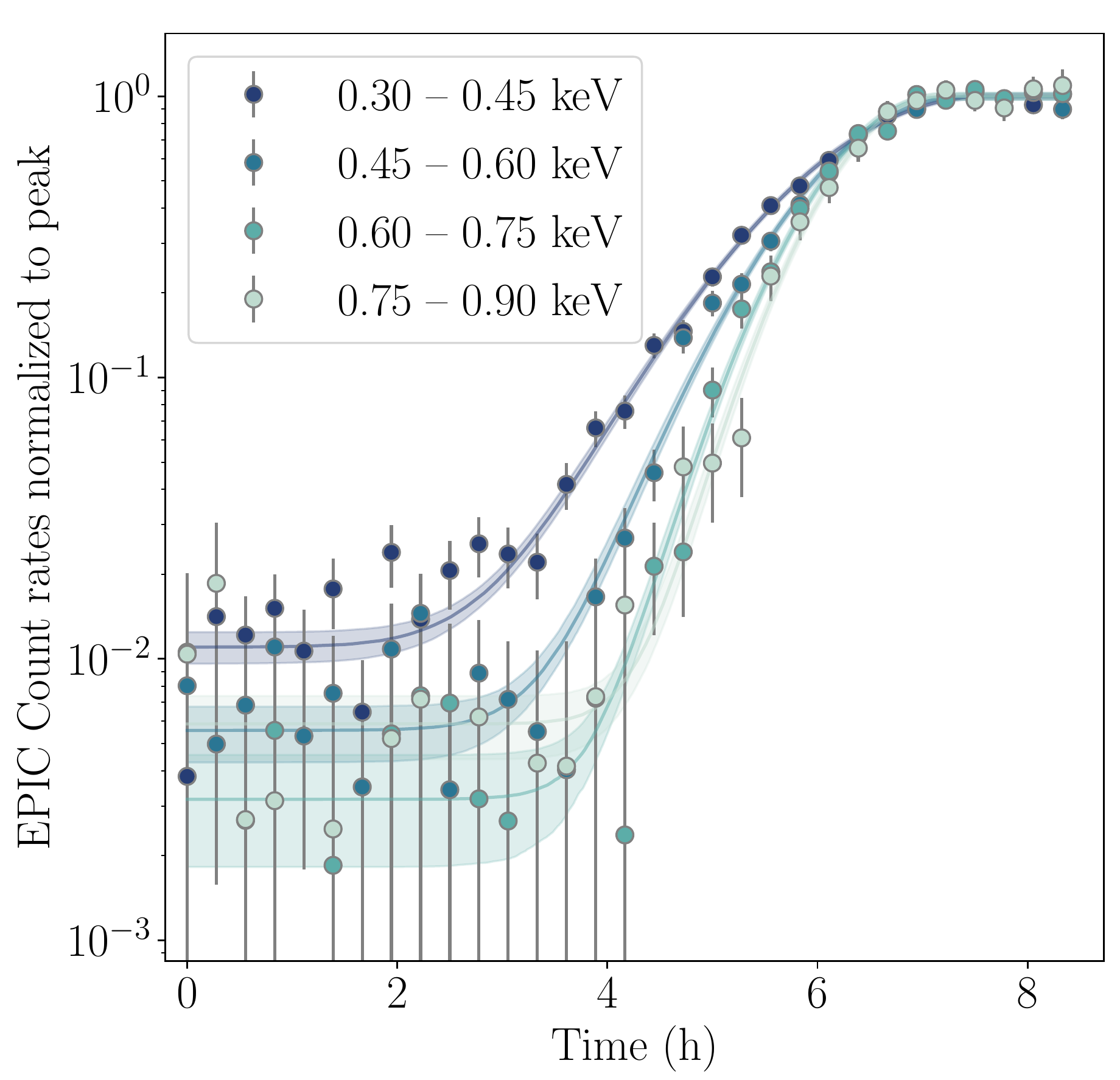}
    \caption{Combined background-subtracted EPIC light curves in different energy bands, binned at 500 s, and normalised to the fitted value of the peaks. The envelopes correspond to the 16$^{\rm th}$ and 84$^{\rm th}$ percentiles of the posteriors, similarly to what is shown in Fig. \ref{fig:light-curvesBands}. This shows the energy dependence of the start of the burst. This is similar to Fig. 2 of \citet{miniutti_nine-hour_2019}, for instance.}
    \label{fig:NormalizedToPeak}
\end{figure}
\end{appendix}

\end{document}